\documentclass[aps,prr,twocolumn,amsmath,amssymb,nobibnotes,longbibliography,superscriptaddress]{revtex4}

\newcommand{\Jnature}{Nature (London)}

\newcommand{\Jprl}{Phys. Rev. Lett.}

\newcommand{\Jpra}{Phys. Rev. A}

\newcommand{\Jpre}{Phys. Rev. E}
\newcommand{\Jrmp}{Rev. Mod. Phys.}

\usepackage{times}

\usepackage[english]{babel}
\usepackage{latexsym}
\usepackage{graphics}
\usepackage{subfigure}
\usepackage{epsfig}
\usepackage{color}
\usepackage{hyperref}
\usepackage{braket} 
\usepackage[T1]{fontenc}
\usepackage{cancel}
\usepackage[latin9]{inputenc}
\usepackage{amstext}
\usepackage{amssymb}
\usepackage{graphicx}
\usepackage{esint}
\usepackage{braket}
\usepackage{babel}
\usepackage{amsmath}

\hypersetup{
colorlinks=true,
citecolor=blue,
linkcolor=red,
urlcolor=black
}


\newcommand{\ie}{i.e.}

\newcommand{\Er}{E_{\textrm{r}}}

\newcommand{\kB}{k_{\textrm{\scriptsize {B}}}}
\newcommand{\fs}{f_{\textrm{s}}}
\newcommand{\ns}{n_{\textrm{s}}}

\newcommand{\rr}{\mathbf{r}}

\newcommand{\asc}{a_{\textrm{\tiny 3D}}}
\newcommand{\aho}{l_\perp}
\newcommand{\atwod}{a_{\textrm{\tiny 2D}}}
\newcommand{\gTwoD}{g_{\textrm{\tiny 2D}}}

\newcommand{\tildeg}{\tilde{g}_{\textrm{\tiny 2D}}}

\newcommand{\fc}{f_\textrm{c}}

\newcommand{\lambdadB}{\lambda_\textrm{T}}

\begin{document}

\title{
Finite size analysis for interacting bosons at the 2D-1D Dimensional Crossover
}

\author{Lorenzo Pizzino}
\affiliation{
DQMP, University of Geneva, 24 Quai Ernest-Ansermet, CH-1211 Geneva, Switzerland
}

\author{Hepeng Yao}\email{Hepeng.Yao@unige.ch}
\affiliation{
DQMP, University of Geneva, 24 Quai Ernest-Ansermet, CH-1211 Geneva, Switzerland
}

\author{Thierry Giamarchi}\email{Thierry.Giamarchi@unige.ch}
\affiliation{
DQMP, University of Geneva, 24 Quai Ernest-Ansermet, CH-1211 Geneva, Switzerland
}

\date{\today}

\begin{abstract}


In this work, we extend the analysis of  interacting bosons at 2D-1D dimensional crossover for finite size and temperature by using field-theory approach ($\emph{bosonization}$) and quantum Monte Carlo simulations. Stemming from the fact that finite size low-dimensional systems are allowed only to have quasi-ordered phase, we consider the self-consistent harmonic approximation to compute the fraction of quasi-condensate and its scaling with the system size. It allows us to understand the important role played by finite size and temperature across the dimensional crossover for deciding the condensate nature. 
Furthermore, we consider a mean-field approximation to compute the finite size effect on the crossover temperature for both weak and strong interaction. All the physical quantities we discuss here provide essential information for quantum gas at dimensional crossover and are directly detectable in cold atom experiments.

\end{abstract}

\maketitle

\section{Introduction}

As a function of dimensionality, quantum systems can show extremely different features: the lower the dimension, the harder it is to have a stable order due to the enhancement of thermal and quantum fluctuations~\cite{bloch-review-2008,giamarchi_book_1d,hadzibabic-2Dgas-2011}. Although most systems are firmly rooted in a well defined dimensionality, systems with strong anisotropies
in one or several directions can exhibit a behavior pertaining to several integer dimensions as a function of a control parameter such as the temperature, or the scale at which the system 
is probed \cite{giamarchi_book_carr,Giamarchi_2004c}. Such a behavior has been observed in several condensed matter setups such as 
weakly coupled fermionic chains~\cite{bourbonnais_review_book_lebed} and weakly coupled spin chains and ladders ~\cite{klanjsek_bpcp,schmidiger_neutrons_bound_spinons,schmidiger_prl2013}.
However in such systems the anisotropy is fixed from the start by the chemistry of the compound, and handles to vary the parameters, such as applying pressure are few and difficult 
to control. 

Given their degree of control on the parameters of the Hamiltonian~\cite{paredes_tonks_experiment,Dalibard-2DBose-2005,Hofferberth2007,bloch-review-2008,meinert-1Dexcitation-2015}
ultracold atoms are thus a perfect setup to study such a phenomenon. The dimensionality of such systems can be continuously controlled by optical lattices or atom chips~\cite{paredes_tonks_experiment,kinoshita_1D_tonks_gas_observation,bouchoule2011,meinert-1Dexcitation-2015}. Recently, various experimental studies have been carried out for bosonic systems at the dimensional crossover, such as correlation evolution~\cite{guo-crossoverD-2023}, anomalous cooling~\cite{guo-cooling-2023}, study of superfluidity~\cite{dalibard2023,spielman-2023,pietro-2024-superfluid}, out-of-equilibrium dynamics~\cite{Hofferberth2007,Jorg-hydrodynamics-crossD-2021} and supersolid phases~\cite{biagioni-supersolid-crossD-2021}. On the theoretical side, coupled bosonic chains have been analyzed for the 3D-1D dimensional crossover~\cite{cazalilla2011,bollmark-crossoverD-2020}. For the 2D-1D crossover, the evolution of the one-body correlation function have been studied in~\cite{yao-crossD-2023,guo-crossoverD-2023}. In addition to the behavior revealed by the 
single particle correlation functions several questions remained to be understood, in particular on the nature of the quasi-long range order through the dimensional crossover, since no true condensate exists in 2D at finite temperature. 

In this work, we thus extend our previous study of the 2D-1D dimensional crossover \cite{yao-crossD-2023} and focus on the (quasi-)condensate evolution by taking into account 
finite size and thermal effects. Combining a field-theory approach (bosonization) with quantum Monte Carlo (QMC) simulations, we study interacting bosons through the 2D-1D crossover with finite size and temperature, with parameters similar to the actual experiments in a cold atom setup. In the strongly-interacting regime, we study the scaling of the quasicondensate fraction $f_c$ as a function of the system size and see how it evolves in the different regimes of dimensionality
and temperatures. This quantity is directly relevant for experiments since it can be directly read from the time-of-flight (TOF) measurement~\cite{plisson-2011,guo-crossoverD-2023}.
We also use a mean-field approach to study the crossover temperature $T_\text{cross}$ from a coherently coupled system of chains to the regime of uncoupled 1D chains, for both weak and strong interaction regimes by recovering the similar scaling factor as predicted in the 3D-1D case~\cite{cazalilla-coupled1D-2006}. We further discuss our findings in connection with the current generation of cold atom experiments, especially comparing with the observed superfluidity~\cite{dalibard2023,spielman-2023} and zero momentum fraction~\cite{guo-crossoverD-2023}.

The paper is structured as follows. In Sec.~\ref{sec:models_and_observables}, we set the playground by presenting the Hamiltonian and the observables. In Sec.~\ref{sec:methods}, we introduce the self-consistent harmonic approximations used to compute the condensate fraction, the mean-field approximation to compute the analytical crossover temperature and quantum Monte Carlo (QMC) details. In Sec.~\ref{sec:results}, we present and discuss the previously introduced observables in the presence of finite size and temperature. In Sec.~\ref{sec:discussion} and \ref{sec:conclusion}, we discuss on possible perspectives and the relevance of our results.

\section{Models and observable}\label{sec:models_and_observables}
In cold atom experiments, the interacting bosonic systems at the 2D-1D crossover can be described by a continuous 2D gas with repulsive two-body contact interactions subjected to a 1D lattice potential $V(\rr)$, with $\rr=(x,y)$ the position of the atom (see Fig.~\ref{fig:sketch_Hamiltonian}).
Its Hamiltonian reads
\begin{equation}\label{eq:Hamiltonian}
  \hat{H} = \sum_j \left[ -\frac{\hbar^2}{2m} \nabla^2_j + V(\hat{\rr}_j) \right] + \sum_{j<k} U(\hat{\rr}_j - \hat{\rr}_k)
\end{equation}
where $\hat{\rr}_j$ is the position of the $j$-th particle and $U$ the short-range repulsive two-body interaction term. The 1D lattice potential along the $y$ direction has a single dependence on the position component $y$ which writes
\begin{equation}\label{eq:QPpotential}
V(\rr) = V_0 \cos^2 (2ky)
\end{equation}
where $V_0$ is the potential amplitude and $k=\pi/a$ is the lattice vectors with $a$ the lattice period.

The potential $U(\hat{\rr}_j - \hat{\rr}_k)$ 
is controlled by the 3D scattering length $\asc$ and the transverse confinement $\aho = \sqrt{\hbar/m\omega_\perp}$~\cite{petrov2000a,petrov-2dscattering-2001}. In practice, this means the 2D coupling constant $\gTwoD$ is written 
\begin{equation}\label{eq:coupling-petrov}
\tildeg \simeq \frac{2\sqrt{2\pi}}{\aho/\asc+{1/\sqrt{2\pi} \ln (1/\pi q^2 \aho^2)}},
\end{equation}
where $\tildeg=m\gTwoD/\hbar^2$ is the rescaled coupling constant and $q=\sqrt{2m|\mu|/\hbar^2}$ is the quasi-momentum.
In the following discussion, we take two examples of the strong and weak interaction regime, namely $\tildeg \simeq 1.36$ and $\tildeg \simeq 0.05$.

\subsection{Bosonization framework}

\begin{figure*}
  \includegraphics[width = 1.75\columnwidth]{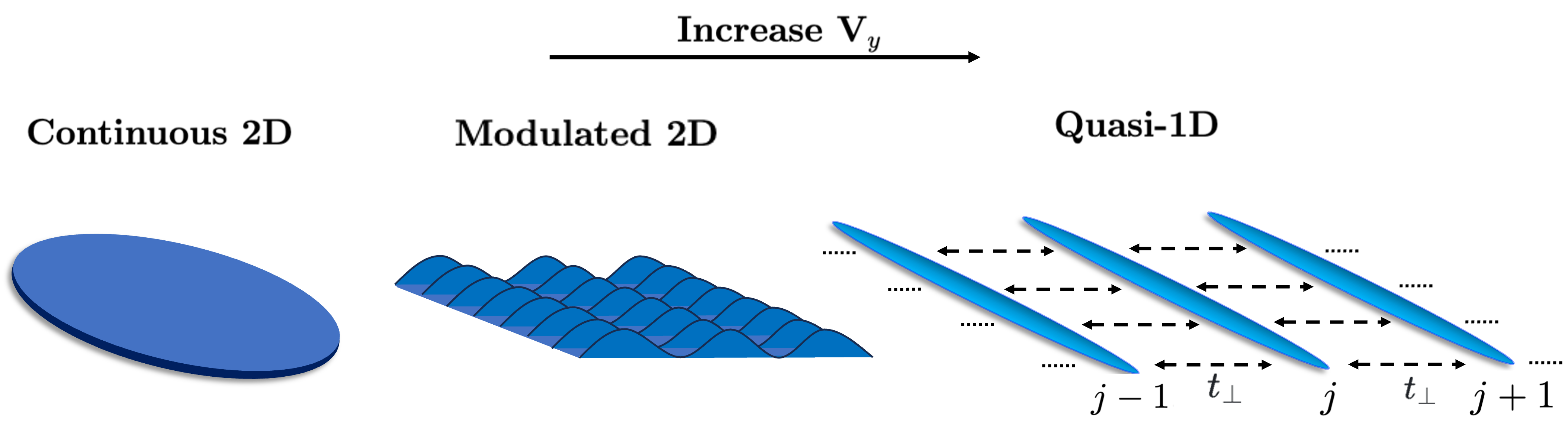}
  \caption{\label{fig:sketch_Hamiltonian}
Sketch of the initial 2D Bose gas undergoing a dimensional crossover as the transverse potential $V(\rr) = V_y = V_0 \cos^2 (ky)$ is increased. The system first evolves from an homogeneous 2D system to a modulated 2D system. For large enough $V_y$, the system can be approximated by an array of continuous 1D chains which are labelled with the letter $j$. The dashed thick lines represent the weak transverse hopping $t_\perp$ that connects different chains in the 2D plane.}
\end{figure*}

As shown in Fig.~\ref{fig:sketch_Hamiltonian}, in the very anisotropic limit (large enough $V_0$ \eqref{eq:QPpotential}), the continuum Hamiltonian can be recast onto a simplified tight-binding model
\begin{equation} \label{H-C1D}
    \begin{aligned}
        \mathcal{H} &= \mathcal{H}_0 + \mathcal{H}_1\\
        &= - t_\parallel \sum_{j=1}^N \sum_{\textbf{R}}  \left (\psi^\dagger_{j,\textbf{R}} \psi^{ }_{j, \textbf{R}} + \text{h.c.} \right )  + \\
        &\qquad \qquad  + U \sum_{j=1}^N \sum_{\textbf{R}}  n_{j,\textbf{R}}\left (n_{j,\textbf{R}}-1 \right) \\
        & \qquad \qquad \qquad  - t_\perp \sum_{j=1}^N  \sum_{\langle \textbf{R,\textbf{R'}}\rangle} \left ( \psi^\dagger_{j,\textbf{R}} \psi^{ }_{j, \textbf{R'}} + \text{h.c.} \right )
    \end{aligned}
\end{equation}
with $\mathcal{H}_{0}$ describing the single chains at position $\textbf{R}$ with repulsive interaction $U$. The transverse hopping term $t_\perp$ in $\mathcal{H}_{1}$ describes particles at position $j$ hopping to nearest neighbor  chains $\langle \textbf{R,\textbf{R'}}\rangle$. In the field-theory treatment we express energy and distance in units of $t_\parallel$ and lattice spacing $a$, respectively. The initial Hamiltonian is now approximated by an array of one dimensional (1D) interacting chains, which are described within the universality class of 1D quantum systems called Tomonaga-Luttinger liquids (TLL) \cite{giamarchi_book_1d}. Within this paradigm, we absorb the effects of interactions and thermal fluctuations. The theory assures a description of the low-energy physics by introducing two smooth fields related by a canonical relation $[\phi(x),\nabla \theta(x')] = i \pi \delta(x-x')$ where $\theta(x)$ and $\phi(x)$ represent phase and density modulation respectively. In this phase-density description, we define the bosonic single-particle operator as $\psi^{\dagger}(x)= \sqrt{A_B \rho(x)}e^{i\theta(x)}$
with $A_B$ a non-universal constants (see Sec.~\ref{sec:results_crossover_temperature}
for more details). We then use these fields to \textit{bosonize} the longitudinal component of the Hamiltonian which now reads \cite{haldane1981}
\begin{equation} \label{eq:free_H_bosonized}
    \mathcal{H}_0 = \frac{1}{2\pi} \int dx \, \left [uK(\nabla \theta(x))^2 + \frac{u}{K}(\nabla \phi(x))^2 \right ]
\end{equation}
and is fully described by two parameters $u$ the velocity of density excitations and $K$ the Luttinger parameter which depend on the amplitude and range of action of interactions. In the limit of hard-core bosons (large repulsive interaction), we have $K=1$ and $u=v_F$ with $v_F$ the Fermi velocity \cite{jordan_transformation1928, cazalilla_rpm_bosons}. Lowering the interaction strength, makes $K$ larger up to $K=\infty$ for non-interacting bosons. 
Up to an artificial cut-off $\Lambda$, the Hamiltonian is quadratic with a linear spectrum $\omega=\pm uk$. 

The coupling between chains in terms of the bosonic fields reads
\begin{equation} \label{eq:Hamiltonian_sine_gordon}
    \mathcal{H}_1 = - t_\perp A_B \rho_0 \ \sum_{\langle \textbf{R}, \textbf{R'} \rangle } \int dx \cos( \theta_{\textbf{R}}(x)- \theta_{\textbf{R'}}(x))
\end{equation}
with $\rho_0$ the unperturbed density. Therefore, to study the dimensional crossover we have to deal with a sine-Gordon (sG) like Hamiltonian \cite{Rajaraman_1982}. Therein, the system is treated as continuous along the chains and discrete in the transverse direction. The effect of $t_\perp$ is to the \textit{build up coherence} along the transverse direction by meaning of fixing the phases. 

\subsection{Observables}
In order to fully characterize the dimensional crossover both analytically and numerically we are particularly interested in studying the one-body correlation function 
\begin{equation}\label{eq:one-body_correlation}
    g_1(x, R_j) = \langle \psi(0,0) \psi^\dagger(x,R_j)\rangle
\end{equation}
with $x$ and $R_j$ the position indices along longitudinal and transverse position respectively. This average encodes the onset of order in the phase field $\theta$ in the system. In order to compare with numerical results, we compute analytically the correlation function within the self-consistent harmonic approximation (SCHA), see Section. \ref{sec:methods} for further details.

The one-body correlation function allows us to study the scaling of the condensation fraction
\begin{equation}
    \label{eq:fc_formula}
    f_c = \frac{n(\textbf{k}=\textbf{0})}{\sum_\textbf{k}n(\textbf{k})}
\end{equation}
for $n(\textbf{k})$ the Fourier transform (FT) of $g_1(x,R_j)$. This quantity measures the fraction of particles that condense at zero momentum.

To complete our study, we compute also the crossover temperature as a function of the transverse hopping
\begin{equation}
    T_\text{cross} = A t_\perp^\nu
\end{equation}
which separates the system from being described by decoupled (\emph{incoherent}) or weakly-coupled (\emph{coherent}) 1D tubes. This temperature gives an energy scale below which the interchain coupling cannot be neglected. Indeed, for small temperatures $T < T_\text{cross}$ the system develops a 2D (quasi-)ordered phase at finite temperature due to the finite size.

\section{Methods} \label{sec:methods}
\subsection{Self-consistent harmonic approximation (SCHA)} \label{sec:SCHA}
In order to solve the Hamiltonian \eqref{eq:Hamiltonian_sine_gordon}, we consider a variational approach called the self-consistent harmonic approximation (SCHA) \cite{Ho2004, cazalilla-coupled1D-2006, Feynman_1972}. The approximation is to make a quadratic ansatz for the action
\begin{equation}
    \mathcal{S}_\text{var}[\theta] = \frac{1}{2 \beta L_x L_y} \sum_{\boldsymbol{q},k_\perp} \mathcal{G}(\boldsymbol{q},k_\perp)^{-1} \theta^*(\boldsymbol{q},k_\perp) \theta(\boldsymbol{q},k_\perp)
\end{equation}
for $\textbf{q} = (k,\omega_n/u)$ a vector with momentum $k$ (along the chain)  and $\omega_n$ the Matsubara frequency, while  $k_\perp$ is the momentum in the transverse direction with $u$ being the sound velocity. Because the variational free energy $\mathcal{F}_{\text{var}}'[\mathcal{G}]$ always overestimates the real free energy $\mathcal{F}$, the parameters $\mathcal{G}(\textbf{q},k_\perp)$ is found by solving  the minimization condition $\delta \mathcal{F}'_{\text{var}} /\delta \mathcal{G}=0$ (see Appendix~\ref{App:SCHA_details})
\begin{align} \label{eq:self_cons_SCHA}
         &\mathcal{G}^{-1}(\textbf{q}, k_\perp)= \frac{K}{\pi u}\left (\textbf{q}^2 + v_\perp^2 F(k_\perp)\right) , \\
         &v_\perp^2 = \frac{\pi u}{K} 2t_\perp A_B \rho_0 e^{-\frac{1}{z\beta L_x L_y  } \frac{\pi u}{K} \sum_{\textbf{q}k_\perp} \frac{F(k_\perp)}{\textbf{q}^2 + v_\perp^2 F(k_\perp)} }\nonumber
\end{align}
with $F(k_\perp)=2\left ( 1-\cos(k_\perp) \right )$ for a 2D system and the transverse lattice spacing $a_\perp = 1$. The resulting $v_\perp$ accounts the \textit{coherence} in the transverse direction and is found self-consistently. Therefore, the result is an effective quantum 1D chain where, at low enough temperature, the transverse hopping results is a correction on top of the free particle case. Setting $t_\perp = 0$ or selecting large temperatures $t_\parallel > T > t_\perp$ gives $v_\perp = 0$ meaning that, for the transverse direction, the system is essentially governed by thermal fluctuations: the resulting action $S_\text{var}$ is the same as the one for isolated chains. If the temperature is even higher than the hopping integral along the longitudinal direction $T > t_\parallel$, the system enters a thermal state and the quantum fluctuations are out of the picture. This means that the approximation is valid as long as $t_\perp \ll 1$ but also $T < t_\parallel$ \cite{Giamarchi_2004c}, for $k_b=1$. 

Within the SCHA, the one body correlation \eqref{eq:one-body_correlation} reads
\begin{widetext}
\begin{equation} \label{eq:g1_SCHA}
    \begin{split}
        g^{\text{SCHA}}_1(x, R_j) 
        &= A_B \rho_0 \exp \left \{-\frac{\pi u}{L_x L_y K} \sum_{k>0, k_\perp} \frac{1 - \cos(kx) \cos(k_\perp R_j)}{\sqrt{u^2 k^2 + v_\perp^2 F(k_\perp)}}\coth \left (\frac{\beta}{2}\sqrt{u^2 k^2 + v_\perp^2 F(k_\perp)} \right ) \right \}
    \end{split}
\end{equation}    
\end{widetext}
where we take the single-particle operator to be $\psi^{\dagger}(x, R_j) \sim \sqrt{A_B \rho_0} e^{i\theta(x,R_j)}$. It is important to recall that the y-direction is treated as discrete and is represented by $R_j$ which numbers the chains. In order to obtain this result, we perform the exact sum of Matsubara frequencies  $-\frac{1}{\beta}\sum_{\omega_n}\frac{1}{i\omega_n -C}= \left (\frac{1}{2}+f_B(C)\right)$, where $f_B(x)= \frac{1}{e^{\beta x} - 1}$ is the Bose distribution \cite{mahan2000}. 

The correlation $g_1$ is essential to address the effects of temperature and finite size of the condensate fraction \eqref{eq:fc_formula}. In the results section, we show how the finiteness of the system gives rise to a non-zero condensate fraction by also investigating various values of $V_y$, which is a continuous parameter that experiments nowadays can directly access.

\subsection{Mean-field approximation: sine-Gordon} \label{sec:T_cross}

Alongside with the SCHA, one direct way of solving the Hamiltonian \eqref{H-C1D} is to apply a mean-field (MF) approximation. In principle, by studying the temperature dependence of $v_\perp$ in \eqref{eq:self_cons_SCHA} we should get the same result and indeed, the same intuitive physics picture holds for both SCHA and MF. Here, we explain how to get an analytical solution of the crossover temperature $T_\text{cross}$ from a MF approach which otherwise would have been only numerical within the SCHA.

By directly decoupling the chains in the transverse direction, we replace $\psi^\dagger_{j,\textbf{R}}  =   \langle \psi^{\dagger}_{j, \textbf{R}}\rangle + \delta \psi^\dagger_{j,\textbf{R}}$
 with the condition that $(\delta \psi^\dagger_{j,\textbf{R}})^2\ll 1$. The system is now represented by a sine-Gordon Hamiltonian of the form
\begin{equation} \label{MF_decouplig_tperp_fieldtheory}
    \mathcal{H} = \mathcal{H}_0 - 2t_\perp z \sqrt{A_B \rho_0} \langle \psi^{\dagger}_{j, \textbf{R'}}\rangle \int dx \cos( \theta_{\textbf{R}}(x))
\end{equation}
with $z$ the number of n.n and $\mathcal{H}_0$ defined in \eqref{eq:free_H_bosonized}. The MF approximation reduced the description to one effective chain where the n.n chains act as a bath via the order parameter $\langle \psi^{\dagger}_{j, \textbf{R}}\rangle$.

The crossover temperature $T_{\text{cross}}$ coincides with the mean-field (MF) critical temperature $T^{\text{MF}}_{\text{cross}}$ above which the order parameter $\Psi_c \equiv \langle \psi^\dagger \rangle =0$ and the bath is not acting on the effective chain. In principle, these two temperatures do not coincide \cite{Giamarchi_2004c}, but in the limiting case of hard-core bosons they do. By observing that as we approach $T^{\text{MF}}_{\text{cross}}$ the order parameter approaches zero $\Psi_c (T \rightarrow T^{\text{MF}}_{\text{cross}}) \ll 1$, we perturbatively compute the self-consistent equation \cite{cazalilla-coupled1D-2006}
\begin{equation} \label{eq:T_cross_mean_field_numerical}
	    1 + t_\perp z g_1^{\mathrm{R}}(k=\omega=0,T \simeq T^{\text{MF}}_{\text{cross}})=0\;,
\end{equation}
with $g_1^{\mathrm{R}}$ being the zero component Fourier transform of the retarded correlation function $g_1(x, \tau) =\langle T_\tau \psi^\dagger(x, \tau), \psi(0,0)\rangle $ with $T_\tau$ the imaginary time order. In the thermodynamical limit $L_x \rightarrow \infty$, such averages $\langle \dots \rangle_0$ are well known quantities and can be exactly computed in 1D systems for quadratic Hamiltonians \cite{giamarchi_book_1d}. The crossover temperature now reads \cite{cazalilla-coupled1D-2006}
\begin{equation}\label{scaling_eq_Tc}
            T^{\text{MF},\infty}_{\text{cross}} = A   t_\perp^{\frac{2K}{4K-1}}
\end{equation}
The microscopic parameters are encoded in the pre-factor $A=[\frac{ z\rho_0 }{u} A_B \alpha^{\frac{1}{2K}}  \sin \left ( \frac{\pi}{4K} \right )  \left (\frac{2 \pi }{u }\right)^{\frac{1-4K}{2K}} B^2 \left (\frac{1}{8K}, \frac{4K-1}{4K} \right )  ]^{\frac{2K}{4K-1}}$ with $B(x,y)= \Gamma(x)\Gamma(y)/\Gamma(x+y)$ the Beta function, $A_B$ the field-theory prefactor that depends on the specific microscopic model we choose and $\alpha$ the built-in short distance cut-off. In the exact result, we used the formula $A_B \sim \left ( \frac{\pi}{K }\right )^{1/2K}$ which is exact within $10\%$ \cite{cazalilla2004_bosonizing_cold}. The scaling in the thermodynamic limit, for hard-core bosons ($K=1$) gives for the crossover temperature
$T^{\text{MF},\infty}_{\text{cross}} \propto t_\perp^{0.66}$ while for soft-core bosons ($K \gg 1$) we have $T^{\text{MF},\infty}_{\text{cross}}\propto t_\perp^{0.5}$.
     
\subsection{Quantum Monte Carlo} \label{sec:QMC}

To carry out calculations beyond mean-field approximation, we rely on the $ab\ initio$ quantum Monte Carlo (QMC) calculations. Specifically, we use the path integral Monte Carlo~\cite{ceperley-PIMC-1995} with worm algorithm implementations~\cite{boninsegni-worm-short-2006,boninsegni-worm-long-2006} in continuous space. It allows us to simulate strongly interacting bosons in continuous potentials, which goes beyond mean-field approximation and tight-binding limit. However, due to the high cost of numerical operation time, it only allows us to simulate a system up to hundreds of lattice sizes. Thus the numerical and analytical approaches are complementary.

Similarly as in Refs.~\cite{yao-crossD-2023,guo-crossoverD-2023,guo-cooling-2023},
we can simulate the Hamiltonian \eqref{eq:Hamiltonian} at a given temperature $T$, 2D scattering length $\atwod$ and chemical potential $\mu$. Then, we can compute the particle density $n$, the superfluid fraction $\fs$ in both directions and the momentum distribution $D(k_x,k_y)$. The finite size condensate fraction $\fc$ can then be written 
\begin{equation}
    \label{eq:fc-QMC}
    \fc=\frac{D(0,0)}{\sum_{j,j'} D(j\frac{2\pi}{L_x},j'\frac{2\pi}{L_y})}
\end{equation}
with $j,j'$ spanning all integers. 
In this manuscript, we use the same QMC algorithm as Refs.~\cite{yao-boseglass-2020, gautier-2Dquasicrystal-2021}. More details about the QMC techniques can be found in these references that contain the computation methods for the superfluid fraction~\cite{yao-boseglass-2020} and the correlation function~\cite{yao-boseglass-2020}, as well as the 2D scattering propagator implementations~\cite{gautier-2Dquasicrystal-2021}. Notably, for the QMC calculations in the following, we always use the lattice spacing $a$ and the corresponding recoil energy $\Er = \pi^2\hbar^2/2ma^2$ as the space and energy units.

\section{Results} \label{sec:results}

In this paragraph we discuss the observables which are relevant for experimental realizations, both in ultracold atoms and condensed matter physics. We directly compare the analytical results with QMC simulations.  First, we present the condensate fraction $f_c$ computed from the SCHA and then we turn to the crossover temperature $T_\text{cross}$ using MF. Furthermore, we highlight the importance of these quantities to spot the dimensional crossover in finite size systems.

\subsection{The quasicondensate fraction: general picture} \label{sec:Condensate_fraction_0}

For low dimensional systems, it is know that a finite condensate fraction is forbidden in the thermodynamic limit at finite temperature. Nevertheless, for a finite system size, as in ultracold atomic experiments, the situation is different: we expect to have a \textit{finite size condensate} with a specific scaling with the system size $f_c \sim L_x^{-\alpha}$.
Therefore, in this section, we study such a scaling with the system size $L_x$ for different values of the inverse temperature $\beta$ and transverse hopping $t_\perp$. Hereafter, we consider hard-core bosons with $K=1$.

\begin{figure}
\centering
\includegraphics[width = 0.95\columnwidth]{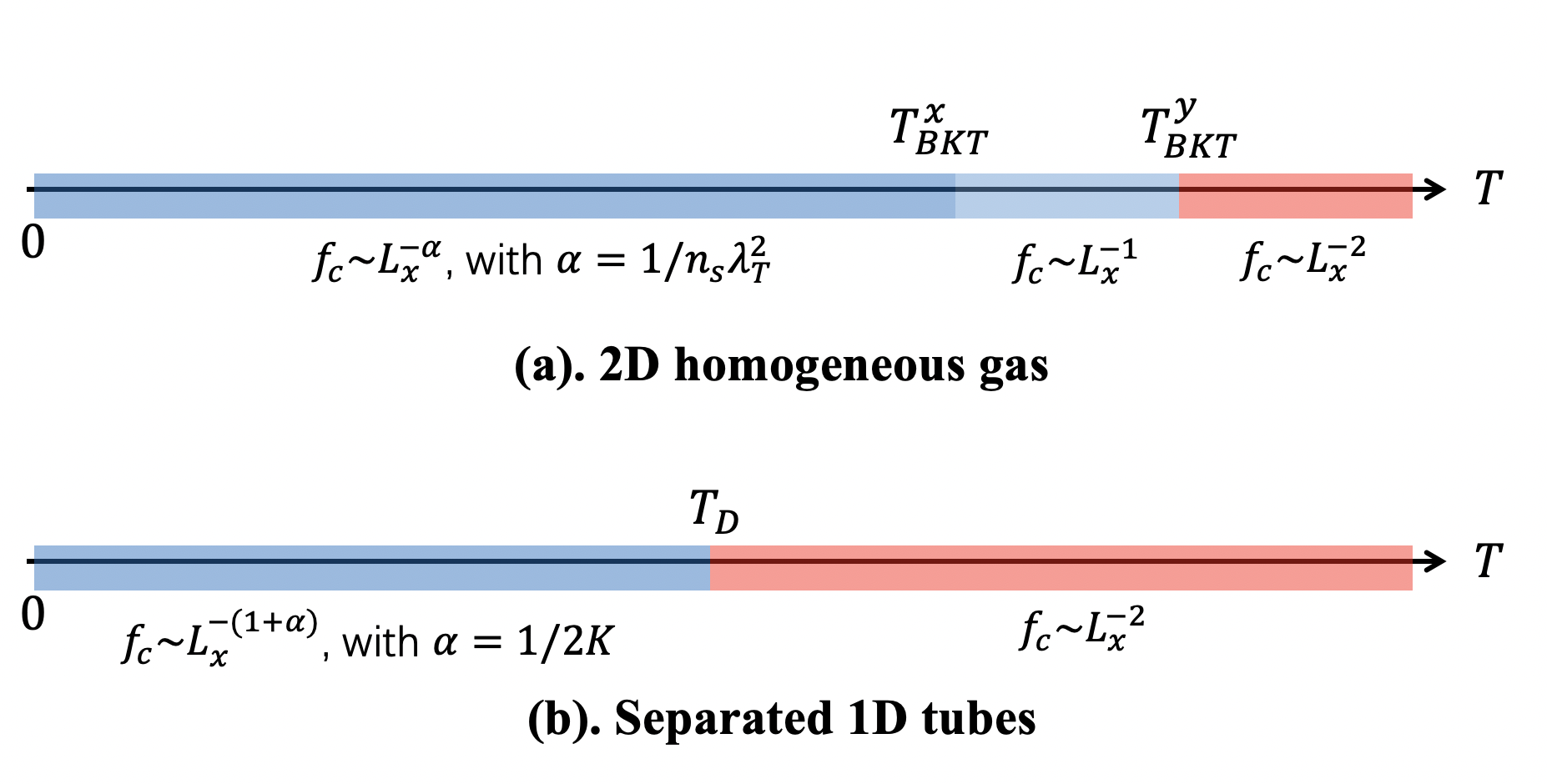}
\caption{\label{fig:fc-L}
The expected scaling, for the two extreme dimensional cases, of the condensate fraction $\fc$ with system size along $x$-direction $L_x$ and the various temperature regimes, with imbalanced system sizes $L_x\gg L_y$ and fixed system size ratio $L_x/L_y=cst$. (a) 2D homogeneous gas, $V_y=0\Er$. (b) Separated-1D tubes, $V_y\gg \Er$. Here, $T^x_{BKT}$ and $T^y_{BKT}$ indicate the BKT transition temperature for $x$ and $y$ directions, correspondingly. 
}
\end{figure}  

Let us first examine general scaling arguments for the condensed fraction. 
Although the finite-size effect can lead to a finite condensate fraction for both 2D and 1D homogeneous systems at finite temperature, the mechanism is quite different. In Fig.~\ref{fig:fc-L}, we demonstrate the expected finite-size scaling of the condensate fraction $\fc$ in the different regimes for homogeneous systems in the two integer dimension limits, with imbalanced system sizes $L_x\gg L_y$ and a fixed ratio $L_x/L_y=cst$. 

In the case of the 2D homogeneous case, the transition to the quasi-ordered regime is of the BKT type, see Fig.~\ref{fig:fc-L}(a). Below a certain temperature $T_{BKT}^x$, the condensate fraction follows $\fc\sim L_x^{-\alpha}$ with $\alpha=1/\ns \lambdadB^2$ the inverse of the quantum degeneracy parameter, see details in Refs.~\cite{bloch-review-2008,hadzibabic-2Dgas-2011}. In the limit of very low temperature $\alpha \ll 1$, we have an almost ordered phase and the correlation can be treated as a constant at large distance, $g_1(\textbf{x} \gg 1)\sim \text{const}$. Due to the anisotropy $L_x>L_y$, the crossover temperature below which such a scaling is valid is different for the two directions $T_{BKT}^x < T_{BKT}^y$.

On the contrary, the 1D gas reaches quantum degeneracy at an even lower degeneracy temperature $T_D<T_{BKT}^x$, see Fig.~\ref{fig:fc-L}(b). For a single 1D tube, the system can be described by a Tomonaga-Luttinger liquid with $\fc\sim L_x^{-\eta}$ with $\eta=1/2K$ when temperature is below $T_D$~\cite{giamarchi_book_1d,cazalilla_review_bosons}. However, since we are considering isolated tubes with $L_y\propto L_x$ on $y$ direction, we thus expect $\fc\sim L_x^{-(1+\eta)}$. By substituting the one-body correlation $g_1$ we recover that
\begin{equation} \label{eq:condensate_fraction_formula}
    \begin{aligned}
        f_c &= \frac{1}{L^2_x}\int d^2\textbf{x} \, \langle \psi(\textbf{0})\psi^\dagger(\textbf{x}) \rangle \propto L_x^{-1-\frac{1}{2K}}
    \end{aligned}
\end{equation}
where $\alpha = 1 + \frac{1}{2K}$ is the expected scaling.

In 1D and 2D, at very high temperature, the system reaches its thermal regime and the one-body correlation $g_1$ decays exponentially at large distances. By substituting in \eqref{eq:condensate_fraction_formula}, we recover that $f_c \sim L^{-2}_x$. 
\begin{figure}
\centering
\includegraphics[width = 0.95\columnwidth]{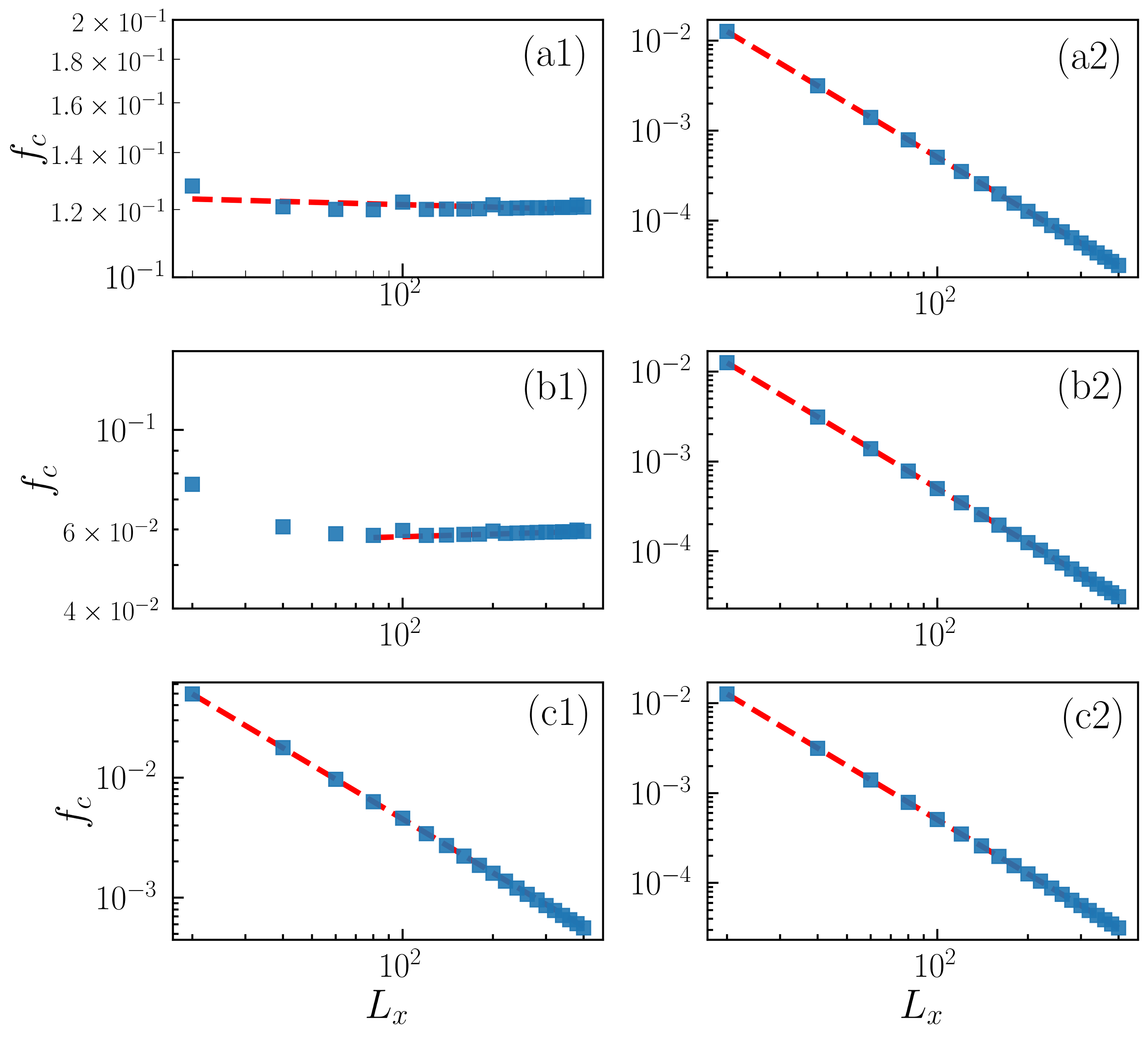}
\caption{\label{fig:analytical_scaling_fc_L}
Analytical finite size scaling for the condensate fraction $\fc=n(k=0)/n$ for $L_x \in [20,400]$  in units of lattice spacing. The four subplots show the condensate fraction $\fc$ as a function of $x$-direction system size $L_x$ in log-log scale for six different cases. a1) $t_\perp = 1$ and $\beta = \infty$, a2) $t_\perp = 1$ and $\beta = 0.1$, b1) $t_\perp = 0.01$ and $\beta = \infty$, b2) $t_\perp = 0.01 $,  $\beta = 0.1$, c1) $t_\perp = 0$ and $\beta = \infty$, c2) $t_\perp = 0$ and $\beta = 0.1$. We remind that we express temperature and hopping in units of $t_\parallel$ for the field-theory treatment.}
\end{figure}

\subsection{The analytical and QMC quasicondensate fraction} \label{sec:Condensate_fraction}

Now, we check this picture with the methods we mentioned in Section \ref{sec:methods}.
In Fig.~\ref{fig:analytical_scaling_fc_L}, we apply the field-theory calculations (blue squares) and check the aforementioned predictions. We consider the situations of strictly-2D $t_\perp=1$ (a1 and a2), coupled-1D $t_\perp=0.1$ (b1 and b2) and strictly-1D $t_\perp=0$ (c1 and c2). In all these cases, we compute the condensate fraction $\fc$ at both low ($\beta = \infty$, left column) and high temperatures ($\beta = 0.1$, right column).

In both the 2D and 1D limits, we find an algebraic decay $\fc\sim L_x^{-\alpha}$ at low and high temperatures. Performing linear fit in log-log scale (red dashed line), we can extract the values of $\alpha$. For a 2D system at low temperature (a1), we find $\alpha_{2D}(T=0)=0.009(4)$, which fits with the predicted value $\alpha=1/\ns \lambdadB^2=0$. Correspondingly, for a 1D system at low temperature (c1), we obtain $\alpha_{1D}(T=0)=1.48(8)$, which also coincides with the prediction $\alpha=1+1/2K=1.5$. Moreover, at high temperature (a2 and c2), we find $\alpha_{2D}(\beta = 0.1)=2.0005(4)$ and $\alpha_{1D}(\beta = 0.1)=2.0005(4)$, and they matches with the expected value $2$. The field theory results are thus confirming the general picture showed in 
Fig.~\ref{fig:fc-L}.

Since the two integer dimension limits $t_\perp=0,1$ behave as predicted, we now use the SCHA for the case of an intermediate value of $t_\perp$ (subplots b). At low temperature, the scaling is clearly not algebraic for the range of system sizes we consider. Clearly, a two-slopes structure appears, which fits with the results in Refs.~\cite{yao-crossD-2023,guo-crossoverD-2023}. It originates from the fact that at the dimensional crossover, the long-range behavior of correlation is still controlled by the higher dimensional properties, while the short-range one already evolves into lower dimensions. Performing algebraic fit on the long-range behavior $L_x \in [100, 400]$, we get $\alpha_{t_\perp=0.1}(T=0)=0.009(0)$ which recovers the 2D scaling at low temperatures.
Furthermore, at the high temperature the decay is still fully controlled by temperature and we recover that $\alpha_{t_\perp=0.1}(\beta = 0.1)=2.0005(4)$, \ie\ the scaling of thermal phase. 

\begin{figure}
\centering
\includegraphics[width = 0.95 \columnwidth]{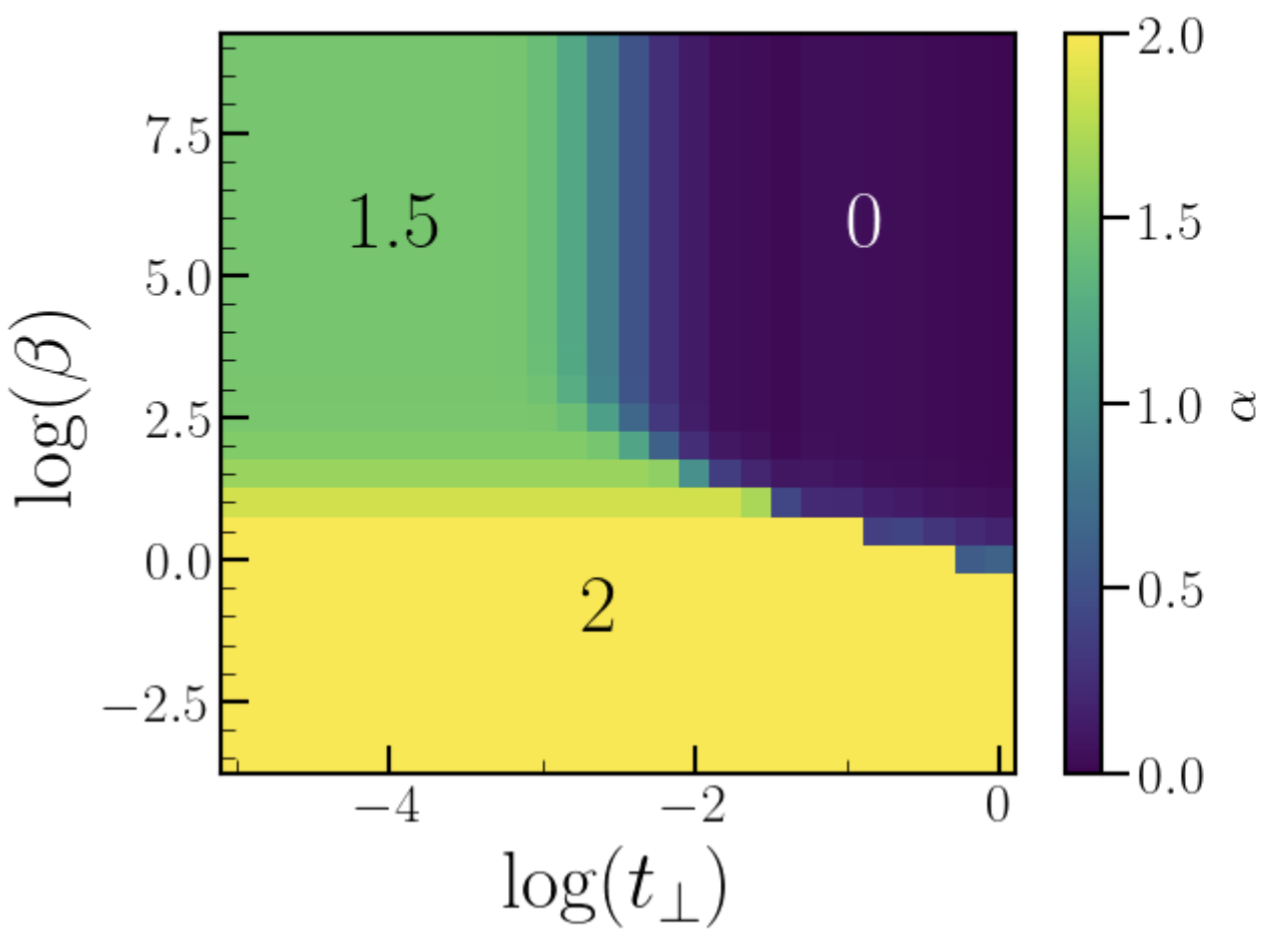}
\caption{\label{fig:analytical_exponent_fc_temperature_t_perp}
Analytical scaling of the finite size condensate fraction $f_c = L_x^{-\alpha}$ as a function of the inverse temperature $\beta$ and transverse hopping $t_\perp$. For large enough temperature, the scaling is $\alpha = 2$ for all values of $t_\perp$, meanwhile at low temperature the exponent evolves from $\alpha=1.5$ to $\alpha=0$.
}
\end{figure}

A more detailed study as a function of $t_\perp$ and temperature $\beta $ is presented Fig. \ref{fig:analytical_exponent_fc_temperature_t_perp}, where we extract the scaling exponent $\alpha$. Here, $\alpha$ is computed by a forced linear fit for the dataset $\log(f_c)$ vs $\log(L_x)$ for $L_x \in [20,400]$. Three distinct plateau regions are presented, namely $\alpha = 2, 1.5$ and $0$. They appear at the bottom, left up corner and right up corner of the diagram, which are high temperature, low temperature 1D and low temperature 2D limits correspondingly. It matches well with our prediction in Fig.~\ref{fig:fc-L}. At the crossover between these regimes, we find properties of dimensional crossover: intermediate values of $\alpha$ appears and it connects the plateaus smoothly.

\begin{figure}
\centering
\includegraphics[width = 0.95\columnwidth]{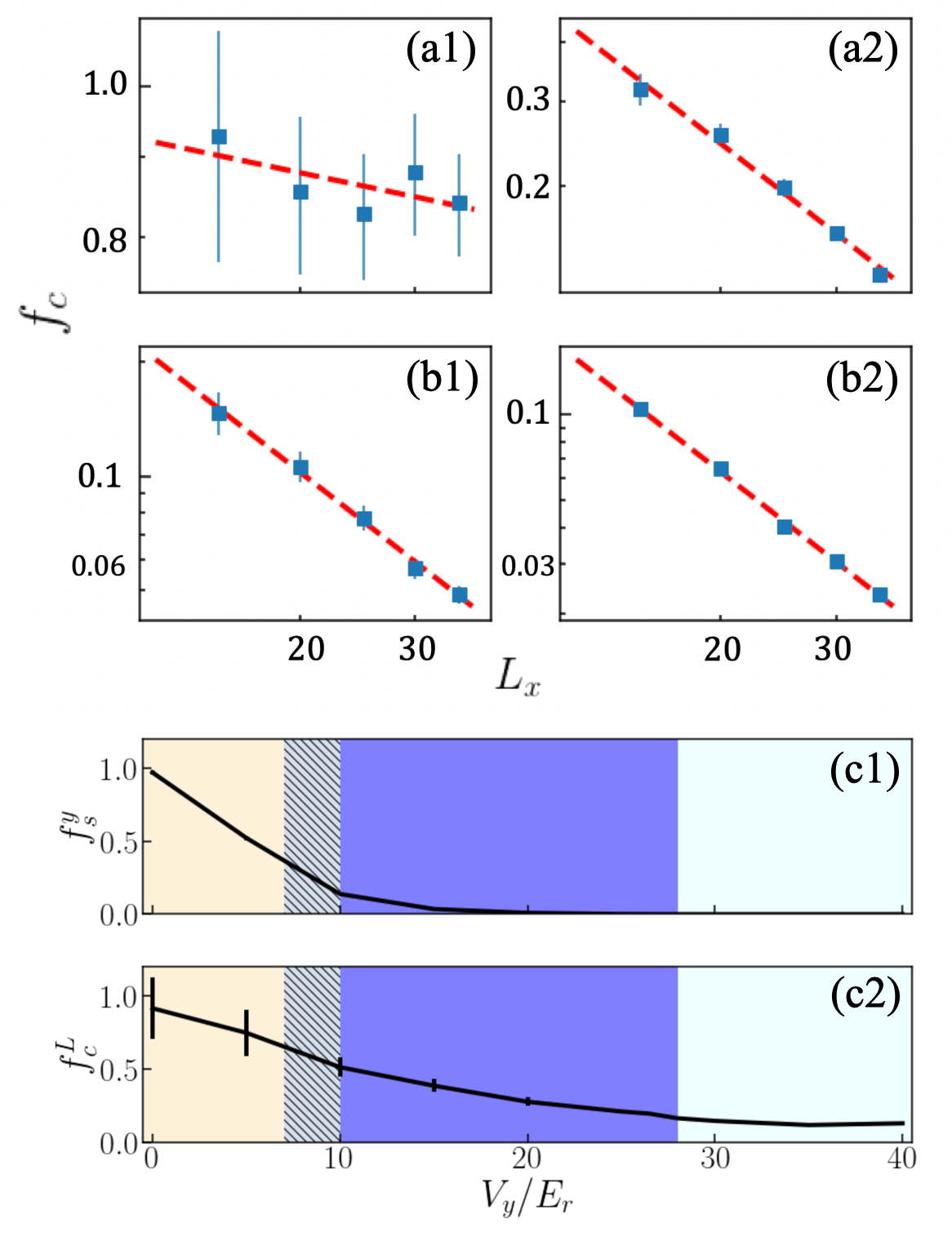}
\caption{\label{fig:fc-data}
The finite size scaling for the condensate fraction $\fc=n(k=0)/n$. The four subplots show the condensate fraction $\fc$ as a function of $x$-direction system size $L_x$ in log-log scale for four different cases, namely: (a1). 2D gas at low temperature, $V_y=0\Er$ and $\beta\Er=74$. (a2). 2D gas at high temperature, $V_y=0\Er$ and $\beta\Er=7$. (b1). 1D gas at low temperature, $V_y=30\Er$ and $\beta\Er=74$. (b2). 1D gas at high temperature, $V_y=30\Er$ and $\beta\Er=18.5$. In all the cases, we fix $L_x/L_y=5$. (c) The quasicondensate fraction as a function
of the y direction lattice depth $V_y$ at fixed temperature $k_B T /E_r = 0.0135$ and system size $L_x=5, L_y=25a$.
}
 \end{figure}

To further confirm the scaling of $\fc$ for such an actual setup, we carry out QMC calculations and compute the size dependence of $\fc$ for the continuous Hamiltonian ~\eqref{eq:Hamiltonian} in various limits. The results are presented in Fig.~\ref{fig:fc-data}. Similarly as for the parameters above, at various temperature $T$ and lattice amplitude $V_y$, we fix the ratio $L_x/L_y=5$ and scan $L_x$. 
In all the cases we considered here, namely 2D and 1D limits at low and high temperatures, we find a scaling $\fc\sim L_x^\alpha$.
In Fig.~\ref{fig:fc-data}(a1), the system is in the 2D regime $V_y=0\Er$ and low temperature limit $\beta\Er=74$. We find $\alpha_{QMC}=0.088\pm0.042$, which fits with the theoretical prediction $\alpha=1/\ns \lambdadB^2=0.1$. 
In Fig.~\ref{fig:fc-data}(a2), the system is in 2D regime $V_y=0\Er$ and high temperature limit $\beta\Er=7$. We find $\alpha_{QMC}=1.06\pm0.06$. Since we find $\fs^x=0$ and $\fs^y>0$, we argue that the temperature is in the regime $T_{BKT}^x<T<T_{BKT}^y$ and thus the $\alpha_{QMC}$ fits with what we expected. At a even higher temperature $\beta\Er=1.4$, the system enters the fully thermal regime and we find $\alpha_{QMC}=1.85\pm0.06$.
Turing to 1D case, in Fig.~\ref{fig:fc-data}(b1), the system is deeply in the 1D regime thanks to the transverse potential $V_y=30\Er$ and low temperature $\beta\Er=74$. We find $\alpha_{QMC}=1.42\pm0.06$, which is $5.5\%$ difference with the theoretical prediction $\alpha=1+1/2K=1.5$ from the Tomonaga-Luttinger theory. 
In Fig.~\ref{fig:fc-data}(b2), the system is in 1D regime $V_y=30\Er$ and high temperature limit $\beta\Er=18.5$. We find $\alpha_{QMC}=1.79\pm0.04$, which is the signature that the system behaves as separated 1D tubes in thermal regime.
Therefore, the QMC results fit with our field theory calculations shown in Fig.~\ref{fig:analytical_scaling_fc_L}. This further confirms the finite-size scaling picture proposed in Fig.~\ref{fig:fc-L}, \ie the quasicondensate nature of finite-size interacting bosons at 2D-1D crossover.


Now, we turn back to the data in Fig.~\ref{fig:fc-data} (a1) and (b1). The system is at temperature $\kB T/\Er=0.0135$ which is below the quantum degeneracy temperatures in both the 2D and 1D regimes for the finite-size system we considered here. Thus, it has a significantly finite quasi-condensate fraction $\fc$ which follows $\fc\sim L_x^{-\alpha}$ with $\alpha_{2D}=1/\ns \lambdadB^2=0.02$ and $\alpha_{1D}=1+1/2K=1.5$ in the 2D and 1D regimes correspondingly. This leads to a much smaller $\fc$ in 1D regimes comparing with the 2D case. This explains the observed significant drop of $\fc$ at this 2D-1D dimensional crossover regime even if no true condensate exists, see Fig.~\ref{fig:fc-data} (c2)~\cite{yao-crossD-2023}.
The value of $\fc$ saturates at a small value in the S1D regime, as presented in Refs.~\cite{bollmark-crossoverD-2020,yao-crossD-2023}. Notably, here one observe a crossover due to the quasi-condensate properties originating from the finite-size effect. This is different from the 3D-1D crossover studied in Ref.~\cite{cazalilla-coupled1D-2006}, where a transition exists even in the thermodynamic limit. Clearly, as shown in Fig.~\ref{fig:fc-data} (c1), a similar signature is observed with the transverse superfluid fraction, which is the experimentally measured quantity in recent works~\cite{dalibard2023,spielman-2023}. These two quantities exhibit behaviors that are qualitatively similar and signal the crossover to 1D regime.

\subsection{Crossover temperature} \label{sec:results_crossover_temperature}

Thanks to the analysis of the condensed fraction discussed above, we find a finite crossover temperature $T_\text{cross}$ from C1D to S1D regime for each given transverse tunneling $t_{\perp}$. In this subsection, we study the properties of this temperature based on various methods, in both the strong and weak interaction regimes.

In the thermodynamic limit, the crossover temperature between coherent and incoherent coupled 1D discrete chains has been already studied and it is predicted to scale algebraically with the coupling between chains \cite{cazalilla-coupled1D-2006}
\begin{equation}\label{fig:Tc-scaling}
    T^{\text{MF},\infty}_{\text{cross}}= A t_\perp^{\nu^{\infty}_{\text{MF}}}
\end{equation}
where $\nu^\infty_{\text{MF}} = \frac{2K}{4K-1}$ is the critical exponent controlled by the Luttinger parameter $K$ and $A$ is the pre-factor. Below this temperature, particles hop in and out the chain coherently. The scaling exponent $\nu$ is also checked by comparing numerically to DMRG simulation of discrete weakly-coupled chains \cite{kantian_crossover_2023, bollmark-crossoverD-2020} and QMC simulations of finite-size continuous systems~\cite{yao-crossD-2023}. It is important to note that as the exponent $\nu$ is an universal feature, while the amplitude $A$ shows different values, see App. \ref{App:pre-factor_bosonsation} for more details.

The above scaling \eqref{fig:Tc-scaling} has been checked numerically for the 3D-1D crossover~\cite{kantian_crossover_2023, bollmark-crossoverD-2020}. However, since no real order exists for 2D systems at finite temperature, its validity for quasicondensate at 2D-1D crossover is not guaranteed. In this section, we show the results of the crossover temperature at $K=1$ and $10$ with two different methods, namely the analytical mean-field crossover temperature $T^{\text{MF}}_{\text{cross}}$ at finite $L_x$ from \eqref{eq:T_cross_mean_field_numerical} and $T^{\text{MF},\infty}_{\text{cross}}$ at infinite $L_x$ from \eqref{scaling_eq_Tc}, and the QMC crossover temperature $T^{\text{QMC}}_{\text{cross}}$ from direct simulations of Hamiltonian \eqref{eq:Hamiltonian}. 
For a quantitative comparison, one needs to make the proper correspondence of the units since the field theory simulates a discrete model with energy unit $t_{\parallel}$ and QMC simulates a continuous system with unit $E_r$.
In Appendix \ref{App:discrete_to_continuous_map} we find that they are linked as $t_\parallel = \hbar^2/2ma^2 =E_r/\pi^2$, with $a$ the space unit and $m$ the particle mass.

\begin{figure*}
    \includegraphics[width = 1.9 \columnwidth]{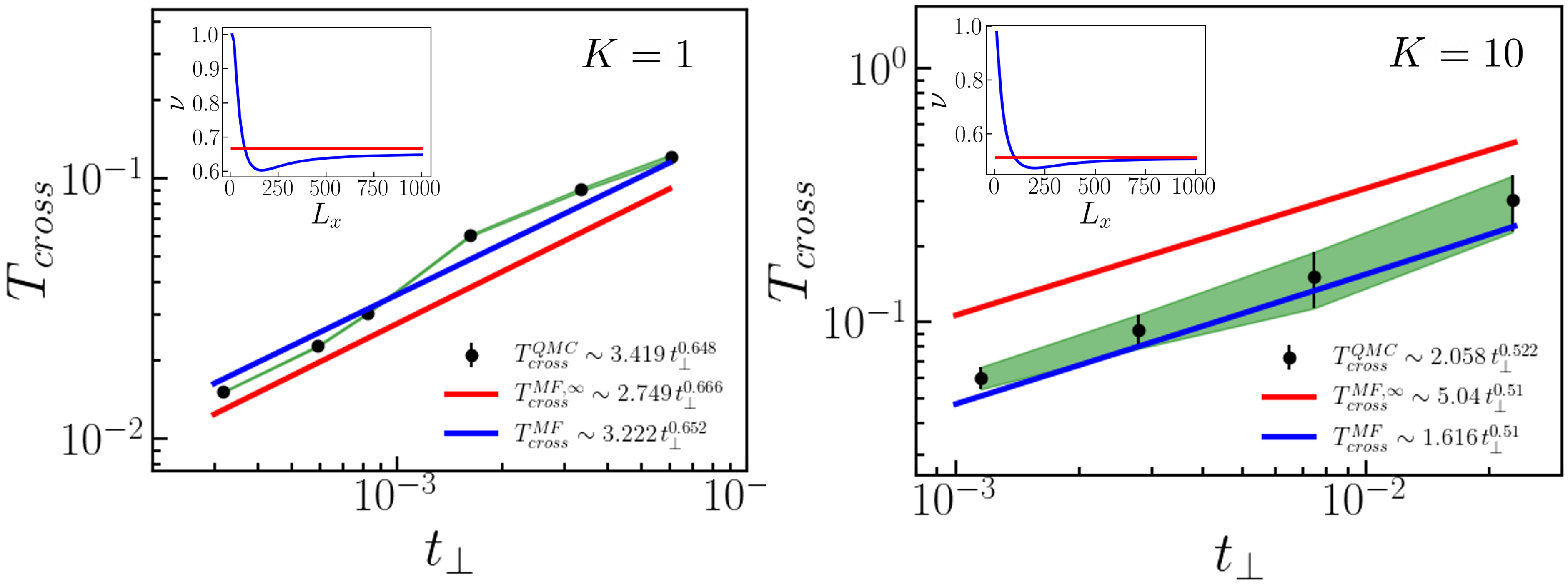}
    \caption{\label{fig:Tc_scaling_tperp_analytics-QMC}
    The comparison of the crossover temperature $T_{\text{cross}}$ as a function of the transverse tunneling $t_{\perp}$ between the analytical (thermodynamic limit and finite system size) and QMC result rescaled accordingly to App. \ref{App:discrete_to_continuous_map}, for the strong interaction $K=1$ and weak interaction $K=10$. We find good agreement in the scaling behavior. The subplot are the scaling exponents $\nu$ of the crossover temperature as a function of the system size compared with the expected analytical values $\nu_{\text{MF}}^{\infty} \sim 0.67$ for $K=1$ and $\nu_{\text{MF}}^{\infty} \sim 0.51$ for $K=10$. The QMC data are for $L_x \times L_y = 25 \times 5$ meanwhile the MF are for  $L_x = 2000$.}
\end{figure*}

For $T^{\text{MF}}_{\text{cross}}$, we consider a large enough system size of $L_x=2000$. The theoretical treatment is limited by the fact that, for a finite system size a too small 
$t_\perp$ would see a gap due to the discretization of the momentum and thus the hoping in the transverse direction would be drastically modified. 
This sets a condition to the minimum system size $L_x>L_{x,{\text{min}}}$, that we can take to have a non-negligible $t_\perp$ effect. For a fixed $t_\perp$, we estimate this requirement by taking the typical scale of the system which gives $L_{x,{\text{min}}}=u/t_\perp$. Indeed, for $L_x<L_{x,{\text{min}}}$ the scaling exponent of $T_{\text{cross}}$ goes up to $\sim 0.8$ in the subplot of Fig.~\ref{fig:Tc_scaling_tperp_analytics-QMC}. This means that for $L_x < L_{x,\text{min}}$ the system behaves as if it was made of uncoupled 1D chains even at zero temperature. It follows that in the thermodynamic limit $L_x \rightarrow \infty$, $L_x>u/t_\perp$ is always satisfied such that any infinitesimally small value of $t_\perp$ is relevant.

In Fig.~\ref{fig:Tc_scaling_tperp_analytics-QMC}, we show the three computed temperatures for the strong interaction case $K=1$. For the scaling exponent, we find an excellent agreement between the field theory calculation  $\nu _{\text{MF}} = 0.652$ with errorbar smaller than $0.1 \%$ and QMC simulation $ \nu_{\text{QMC}} = 0.648\pm 0.02$. The expected analytical result in thermodynamic limit is $\nu^\infty_{\text{MF}} = \frac{2K}{4K-1}\sim 0.66$ (see the red solid line in subplot of Fig. \ref{fig:Tc_scaling_tperp_analytics-QMC}). These values agree with each other within $10\%$. Even for the prefactor $A$, they only exhibit a difference of $5\%$. Therefore, the field theory calculation gives similar results for the crossover temperature as the one for the continuous systems. This also proves that the analytical expression for the scaling factor in Eq.~\ref{fig:Tc-scaling} still holds for 2D-1D crossover.

We now turn to the weak interaction case which results in a larger value of the TLL parameter, namely $K=10$. As we did for $K=1$, in Fig. \ref{fig:Tc_scaling_tperp_analytics-QMC}, we perform a linear fit in log-log scale and find the scaling exponent $\nu _{\text{MF}} = 0.51$ with errorbar smaller than $0.1 \%$ and QMC simulation $ \nu_{\text{QMC}} = 0.52 \pm 0.05$. They agree well with the expected exponent of $\nu^\infty_{\text{MF}} = \frac{2K}{4K-1}\sim 0.51$ within $5\%$. However, a strong deviation exists for the prefactor $A$, where a $20\%$ difference appears. We argue that it is due to the fact that the field theory description is less valid when $K$ is large. Indeed, lowering the interactions reduces the value of the chemical potential and therefore limits the validity of the low-energy description.

\section{Discussion} \label{sec:discussion}

Previous works~\cite{cazalilla_06,bollmark-crossoverD-2020}, mainly focused on the 3D-1D crossover where a phase transition between the real ordered phase (true condensate) and incoherently-coupled 1D tubes appears.
Here, we rather focus on the 2D-1D dimensional crossover where no true condensate exists in the thermodynamic limit. This suggests that finite-size effect can play an important role and a transition linked with the quasicondensate properties takes place, which further complicates the description. By carefully carrying out the study of finite-size effect, we find the coherence of the system is determined by the scaling exponent which depends on temperature or interaction. If it is small enough such that the correlation decay can be neglected up to the system size, a quasicondensate is formed. Thus, the nature of 2D-1D dimensional crossover for finite-size systems can be clearly reflected by the properties of the exponent $\alpha$, \ie\  Fig.~\ref{fig:analytical_exponent_fc_temperature_t_perp}.

In real cold atom experiments, the dimensional crossover has been measured by one-body correlation functions~\cite{guo-crossoverD-2023} and superfluid fraction~\cite{dalibard2023,spielman-2023}. Here, we prove that the condensate fraction $\fc$ also carries important information for dimensional crossover. Notably, this is a quantity that can be easily obtained from the momentum distribution, which can be directly detected via time-of-flight techniques. Such measurement is widely used in cold atom experiments, and it is less complex comparing with others. Therefore, our work provides a new path to quickly access the coherence properties of cold atom systems at dimensional crossover.

\section{Conclusion} \label{sec:conclusion}

In this work, we study the finite size effects for interacting bosons at the continuous 2D-1D crossover. We first investigate the quasicondensate fraction $f_c$ scaling with the system size $L_x$ for different fixed temperatures and transverse hopping $t_\perp$ in the limit of strong interaction $K=1$. We illustrate the nature of quasicondensate according to the finite-size scaling. We find that the decay exponent evolves from $\alpha_{2D}=0$ to $\alpha_{1D}=1.5$ at low temperature which signals the crossover from a 2D to a 1D quasicondensate. We also find that this exponent evolves to $\alpha_{\mathrm{TH}}=2$ at high temperature, where the system is in a thermal phase. In the second part, we turn to the crossover temperature $T_{\text{cross}}$ both for the strong interaction $K=1$ and the weak interaction $K=10$ limit.  The crossover temperature we find between the 2D and 1D quasicondensate, follows a similar scaling as the one of 3D-1D, where a true condensate exists at the higher dimensionality.

Our results prompts for further theoretical investigation of higher order correlation functions~\cite{Cheneau2012,spielman-2023} and momentum space correlation~\cite{Carcy-2019-kcorrelation,tenart-2021-kcorrelation} at the dimensional crossover. It would also be interesting to investigate the finite size effect and zero momentum fraction for 3D-1D and 3D-2D dimensional crossover, since there exists a true BEC in 3D. Recent experiments on 2D weakly-interacting bosonic gases with transverse potential $V_y$ ~\cite{dalibard2023,spielman-2023} have shown interesting comparison to the Leggett's bounds \cite{leggett1970,leggett1998superfluid} of the transverse superfluidity $f_s^y$. In principle, the $f_s^y$-$V_y$ exhibits similar properties as the $f_c^L$-$V_y$ one (Fig. 5(e)) in the low temperature regimes, which indicates the crossover to 1D system. The validity and usefulness of these bounds for strong-interactions is still under debate and needs further investigations. 

\acknowledgments
This work is supported by the Swiss National Science Foundation under grant number 200020-188687 and 200020-219400. Numerical calculations make use of the ALPS scheduler library and statistical analysis tools~\cite{troyer1998,ALPS2007,ALPS2011}. We would like to thank S. Stringari, J. Dalibard, P. Massignan and L. Sanchez-Palencia for interesting discussions.  
\appendix

\section{SCHA detailed calculations}
\subsection{SCHA action} \label{App:SCHA_details}
In this section we show how to implement the SCHA.
We start by making the variational Ansatz of an action quadratic in the phase field
\begin{equation}
    \mathcal{S}_\text{var}[\theta] = \frac{1}{2 \beta L_x L_y} \sum_{\textbf{Q}} \mathcal{G}(\textbf{Q})^{-1} \theta^*(\textbf{Q}) \theta(\textbf{Q})
\end{equation}
with $\textbf{Q}=(\textbf{q},\textbf{k}_\perp)$. Here, $\mathcal{G}(\textbf{Q})$ is a set of parameters that, at this stage, we still have to fix. By using the properties of Gaussian integration
\begin{equation}  \label{gaussian_average}
    \begin{aligned}
    \langle u^*_{\textbf{q}_1} u^{}_{\textbf{q}_2} \rangle &= \frac{\int \mathcal{D}u[\textbf{q}] \, u^*_{\textbf{q}_1} u^{}_{\textbf{q}_2}  e^{-\frac{1}{2}\sum_\textbf{q}u^*_{\textbf{q}}S(\textbf{q})u^{}_{\textbf{q}}}}{\int \mathcal{D}u[\textbf{q}] e^{-\frac{1}{2}\sum_\textbf{q}u^*_{\textbf{q}}S(\textbf{q})u^{}_{\textbf{q}}}} \\
    &= \delta_{\textbf{q}_1, \textbf{q}_2} \frac{1}{S(\textbf{q}_1)}
    \end{aligned}
\end{equation}
we compute the variational free energy. First we rewrite the partition function as
\begin{equation}
    \mathcal{Z} = \int\mathcal{D}\theta \, e^{-\mathcal{S}}=\mathcal{Z}_{\text{var}} \big \langle e^{-( \mathcal{S}-\mathcal{S}_\text{var})} \big \rangle_{\text{var}}
\end{equation}
Then, we substitute and use the convexity condition $(k_b=1)$
\begin{equation}
    \begin{aligned}
        \mathcal{F} \leq \mathcal{F}_{\text{var}}'[\mathcal{G}] &= - T\ln\left (\mathcal{Z}_{\text{var}} \big \langle e^{-( \mathcal{S}-\mathcal{S}_\text{var})} \big \rangle_\text{var} \right ) \\
        &= \mathcal{F}_{\text{var}} + T \big \langle  \mathcal{S}-\mathcal{S}_\text{var} \big \rangle_\text{var}
    \end{aligned}
\end{equation}
From the above equation, we conclude that the variational free energy always overestimates the real free energy. The first term gives $\mathcal{F}_{\text{var}}  = -   T \sum_{\boldsymbol{Q} > 0} \, \ln \mathcal{G}_{\boldsymbol{q}} + \text{const}$.
For the second term we observe that $\langle\mathcal{S}_{\text{var}}\rangle_{\text{var}}$ is a constant and therefore we compute now
\begin{widetext}
\begin{equation}
    \begin{aligned}
        \langle  \mathcal{S} \big \rangle_\text{var}
        & = \frac{1}{\mathcal{Z}_\text{var}[\theta]}\iint\mathcal{D}\theta_\textbf{Q} \mathcal{D}\theta^*_\textbf{Q} \, e^{-\mathcal{S}_\text{var}[\theta]} \Bigg [ \sum_{\textbf{R}}\frac{K}{2\pi} \int_0^{L_x} dx \int_0^\beta  d\tau \, \left ( \frac{1}{  u}(\partial_{\tau}\theta_{\textbf{R}})^2 +   u(\partial_x \theta_{\textbf{R}})^2 \right) + \\
        & \qquad \qquad \qquad \qquad - t_\perp A_B \rho_0  \sum_{\left <\textbf{R,R'} \right>} \int_0^{L_x} dx \int_0^\beta  d\tau \, \cos \left (\theta_\textbf{R}(x,\tau) - \theta_{\textbf{R}'}(x,\tau) \right) \Bigg ] + \text{const}\\
        &= \frac{1}{2} \frac{1}{ \beta L_x L_y}\sum_{\textbf{Q}}\frac{K}{\pi u} \left ( \omega_n^2 +   u^2 k^2 \right)\frac{1}{\mathcal{Z}_\text{var}[\theta]}\iint\mathcal{D}\theta_\textbf{Q} \mathcal{D}\theta^*_\textbf{Q} e^{-\frac{1}{2} \sum_{\textbf{Q}} \big ( \frac{\mathcal{G}_\text{var}^{-1}}{\beta L_xL_y } \big ) \theta^*_\textbf{Q}\theta^{}_\textbf{Q}}\theta^*_{\boldsymbol{Q}} \theta^{}_{\boldsymbol{Q}} - t_\perp A_B \rho_0 \dots \\
        &= \frac{1}{2} \sum_{\boldsymbol{Q}} \mathcal{G}_0^{-1}(\textbf{q})  \mathcal{G}(\textbf{Q}) - t_\perp A_B \rho_0 \int_0^{L_x} dx \int_0^\beta d\tau \,  \sum_{\left <\textbf{R,R'} \right>} \frac{2 \, e^{-\frac{1}{2}\langle\left [\theta_\textbf{R}(0,0) - \theta_{\textbf{R'}}(0,0) \right]^2 \rangle_\text{v}}}{2}  \\
    \end{aligned} 
\end{equation}
\end{widetext}
with $ \mathcal{G}_0^{-1}(\textbf{q}) = \frac{K}{\pi u} \left ( \omega_n^2 +   u^2 k^2 \right) $.
In the second last step, we used the fact that we have a quadratic action to substitute $\langle e^{i\theta} \rangle = e^{-\frac{1}{2}\langle \theta^2\rangle}$.
After neglecting the constant term, in the first step, we Fourier transform the quadratic action, which gives the factor $\frac{1}{\beta L_xL_y}$. Then we regroup the terms and insert the Dirac delta. At this point, we are left with the average $\langle \theta^*(\textbf{Q}) \theta(\textbf{Q})\rangle$. The Fourier transform and the definition of Dirac delta we use are
\begin{equation}
    \begin{aligned}
        &\theta_{\textbf{R}}(x,\tau) = \frac{1}{\beta L_x L_y} \sum_{\textbf{q}, \textbf{k}_\perp}e^{i(\textbf{q}\cdot \textbf{r} + \textbf{k}_\perp \cdot \textbf{R})}\theta(\textbf{q}, \textbf{k}_\perp) \\ &\delta_{\textbf{q},-\textbf{q'}} \delta_{\textbf{k}_\perp, -\textbf{k'}_\perp} = \frac{1}{\beta L_x} \int dx d\tau \, e^{i(\textbf{q}+\textbf{q'})\cdot \textbf{r}}\frac{1}{L_y} \sum_{\textbf{R}} \, e^{i(\textbf{k}_\perp+\textbf{k'}_\perp)\cdot \textbf{R}} 
    \end{aligned}
\end{equation}
where $\textbf{q}\cdot \textbf{r}= kx - \omega_n\tau$. Then we do the Gaussian integral which gives $\mathcal{G}(\textbf{Q})$. In the third step, we use the symmetry under translation to substitute $(x,\tau) \rightarrow (0,0)$ inside the average. The last term is now of the form
\begin{equation}
    \begin{aligned}
        \langle \theta_{\textbf{R}'}(\textbf{0})\theta_{\textbf{R}}(\textbf{0})\rangle 
        &= \frac{1}{\beta L_x L_y} \sum_{\textbf{Q}} e^{i\textbf{k}_\perp \cdot ( \textbf{R} - \textbf{R}')} \mathcal{G}(\textbf{Q})\\
    \end{aligned}
\end{equation}
where we insert the Gaussian average that $\langle \theta_{\mathbf{Q}} \theta _{\mathbf{Q}'}\rangle = \frac{1}{\mathcal{Z}[\theta]}\int \mathcal{D}\theta[{\mathbf{Q}}] \, \theta_{\textbf{Q}} \theta_{\textbf{Q'}} \, e^{-\frac{1}{2}\sum_\textbf{Q}\theta^*_\textbf{Q}\mathcal{A}^{ }_\textbf{Q}\theta^{ }_\textbf{Q}} = \delta_{\textbf{Q}, -\textbf{Q'}} \mathcal{A}^{-1}_\textbf{Q}$. By defining the Fourier transform
\begin{equation}
    \mathcal{G}(x, \textbf{R-R'}, \tau) = \frac{1}{\beta L_xL_y} \sum_{\textbf{Q}} e^{i\textbf{k}_\perp \cdot (\textbf{R-R'}) + ikx-i\omega_n\tau} \mathcal{G}(\textbf{Q})
\end{equation}
we are set to rewrite the exponent
    \begin{equation}
    \begin{aligned}
        \mathcal{G}(0,\textbf{R-R'},0) - \mathcal{G}(0,\textbf{0},0) =-\frac{1}{z}\frac{1}{\beta L_xL_y} \sum_{\textbf{Q}}  F(\textbf{k}_\perp) \mathcal{G}(\textbf{Q})
    \end{aligned}
\end{equation}
with $F(\textbf{k}_\perp) = \sum_{ \textbf{R-R'}} \Big ( 1-\cos({\textbf{k}_\perp \cdot (\textbf{R-R'})}) \Big )$ where we use that $1=\frac{1}{z}\sum_{ \textbf{R-R'}}$.
We now minimize the variational free-energy:
\begin{widetext}
    \begin{equation}
    \begin{aligned}
        \frac{\delta \mathcal{F}'\left[\mathcal{G}\right ]}{\delta \mathcal{G} } &=\frac{\delta}{\delta \mathcal{G} } \Bigg[ - \frac{1}{\beta} \sum_{\textbf{Q}>0} \ln \mathcal{G}(\textbf{Q})  + \frac{1}{2\beta}\sum_{\boldsymbol{Q}} \mathcal{G}_0^{-1}(\textbf{q})  \mathcal{G}(\textbf{Q}) - \frac{1}{\beta} t_\perp A_B \rho_0 \beta L_x \sum_{\left \langle \textbf{R,R'} \right \rangle} e^{-\frac{1}{z \beta L_xL_y} \sum_{\textbf{Q}}  F(\textbf{k}_\perp) \mathcal{G}(\textbf{Q}) }  + \text{const}  \Bigg ] \\
        &= \frac{1}{\beta} \sum_{\boldsymbol{Q}>0} \left[ - \frac{1}{\mathcal{G}(\textbf{Q})}+  \frac{1}{\mathcal{G}_0(\textbf{q})}   +  2t_\perp A_B \rho_0 F(\textbf{k}_\perp) e^{-\frac{1}{z \beta L_xL_y} \sum_{\textbf{Q}}  F(\textbf{k}_\perp) \mathcal{G}(\textbf{Q}) }  + \text{const}  \right ]
    \end{aligned}
\end{equation}
\end{widetext}
where we use that the sum over n.n. gives $\sum_{\langle \textbf{R}, \textbf{R'}\rangle} 1= \frac{L_y z}{2}$ and that and $F(\textbf{k}_\perp) = F(-\textbf{k}_\perp)$. By imposing $\frac{\delta \mathcal{F}'\left[\mathcal{G}\right ]}{\delta \mathcal{G} } = 0$, the \emph{optimal} set of parameters reads
\begin{equation}\label{SCHA_self_cons_equation}
    \frac{1}{\mathcal{G}(\textbf{Q})} = \frac{1}{\mathcal{G}_0(\textbf{q})} + 2t_\perp A_B \rho_0 F(\textbf{k}_\perp) e^{-\frac{1}{z\beta L_xL_y} \sum_{\textbf{Q}}  F(\textbf{k}_\perp) \mathcal{G}(\textbf{Q}) } 
\end{equation}
and has to be found self-consistently. In the limiting case of $t_\perp=0$, we have $\mathcal{G}(\textbf{Q}) = \mathcal{G}_0(\textbf{q})$. Moreover, we observe how this approximation allows us to keep track of the system geometric as well as the system size in all spatial directions. 

\subsection{SCHA one-body correlation function}
Once we have the quadratic action we are able to compute the one-body correlation function $g^{\text{SCHA}}_1(x, j) = \langle \psi(0,0) \psi^\dagger(x,j)\rangle = A_B \rho_0 e^{-\frac{1}{2} \langle[\theta(0,0) - \theta(x,j) ]^2\rangle_{\text{SCHA}}} $ where we use the single-particle operator to be $\psi(x,j) = \sqrt{\rho(x,j)} e^{i\theta(x,j)}$.
By substituting the Fourier transform of the phase field $\theta_{\textbf{R}}(x, \tau)= \frac{1}{\beta L_xL_y} \sum_{\textbf{Q}}  \theta_\textbf{Q}e^{i(\textbf{q}\cdot \textbf{r}+\textbf{k}_\perp \cdot \textbf{R})}$
the average reads
\begin{widetext}
    \begin{equation} \label{correlation_scha_analytica_zero_finite_temp_sum_matsubara}
    \begin{aligned}
        \langle [\theta_{\textbf{R}}(x, \tau)- \theta_{\textbf{0}}(0,0)]^2 \rangle_{\text{SCHA}} &= \frac{1}{(\beta L_xL_y)^2} \sum_{\textbf{Q},\textbf{Q}'} \langle \theta_\textbf{Q} \theta_{\textbf{Q}'}\rangle_{\text{SCHA}}\Big [  e^{i(\textbf{q}\cdot \textbf{r}+\textbf{k}_\perp \cdot \textbf{R})} - 1 \Big ] \Big [  e^{i(\textbf{q}^\prime\cdot \textbf{r}+\textbf{k}_\perp^\prime \cdot \textbf{R})} - 1 \Big ]\\
        &= \frac{1}{\beta L_xL_y K} \sum_{\textbf{Q}} \frac{2 \pi u}{\omega_n^2 + u^2 k^2 + v_\perp^2 F(\textbf{k}_\perp)}\Big [1 - \cos(kx-\omega_n\tau+\textbf{k}_\perp \cdot \textbf{R}) \Big ] \\
        & \stackrel{\tau = 0}{=}  \frac{2\pi u}{ L_xL_y K} \sum_{k>0, \textbf{k}_\perp} \frac{  \coth \left (\frac{\beta}{2}\sqrt{u^2 k^2 + v_\perp^2 F(\textbf{k}_\perp)} \right ) }{\sqrt{u^2 k^2 + v_\perp^2 F(\textbf{k}_\perp)}}\Big [1 - \cos(kx) \cos(\textbf{k}_\perp \cdot \textbf{R}) \Big ]
    \end{aligned}
\end{equation}
\end{widetext}
In the last line, we perform the exact sum over the Matsubara frequencies  at the equal imaginary time ($\tau =0)$ because the system is Galilean invariant and $r^2 = x^2 + u^2\tau^2$.
In the limit of zero temperature the formula simplifies because $\coth ( \frac{\beta}{2} \dots ) \rightarrow 1$. We emphasize that for the (\emph{bosonized}) longitudinal direction we sum up to an artificial cut-off $\Lambda$ (e.g. $\Lambda = \frac{\pi}{2 a}$, with $a$ the lattice spacing). It is similar to say that theory describes spatial fluctuations larger than a certain length $1/\Lambda$. For the transverse direction we impose periodic boundary conditions.

\section{Fixing the pre-factor of the single-particle operator}
\label{App:pre-factor_bosonsation}
In order to be quantitative, it is important to treat carefully the prefactor of $T_\text{cross}$ from analytics. Not only it depends on the interactions and filling as the TLL parameter $K$ does but also on the other parameters of the system. For this reason, it is worth to understand better how to fix the non-universal prefactor $A_B$ hidden in the pre-factor $A$, which defines the single-particle creation operator in the field-theory. For this, it is necessary to correctly map the field-theory to the specific microscopic model used in QMC simulation. Given that the field-theory depends on a cut-off $\Lambda$, then $A_B$ has to scale properly such that for medium distances $x>1/\Lambda$ correlation $g_1(x)$ from QMC and field-theory match. By comparing the results, we conclude that in the limit of strong interaction $A_B(K=1) \sim 2.15$ and for weak interaction is $A_B(K=10) \sim 1.13$.

\section{Discrete to continuous mapping} \label{App:discrete_to_continuous_map}
Here, we explicitly show how to link the longitudinal hopping $t_\parallel$ of the tight-binding model (starting point of field-theory treatment) with the recoil energy $E_r$ of the continuous limit (simulated by QMC). In other words, we want to compare the energy scale in the continuous limit, where $H = - \frac{\hbar^2}{2m}\nabla^2 \psi$, and in the discrete limit where we start from a tight-binding model $\hat{H} = -t_\parallel \sum_j \left (\hat{a}^\dagger_j \hat{a}^{}_{j+1} + \hat{a}^\dagger_j \hat{a}^{}_{j-1}\right )$.

For the continuous case, if we call $\Delta x$ the smallest spatial variation we can describe, then the derivative can be recast in
\begin{equation}
    \partial_x \psi_j = \frac{\psi_{j+1}-\psi_{j}}{ \Delta x}
\end{equation}
with $\psi_j$ the wave-function at position $j$. Therefore, the eigenvalue equation is of the form 
\begin{equation} \label{eq:continuum_eigen}
    E \psi_j = -\frac{\hbar^2}{2m (\Delta x)^2}\left ( \psi_{j+1} + \psi_{j-1}\right )
\end{equation}
On the other hand, for the discrete limit we start by explicitly define the wave-function at position $j$ as $\langle j|\psi\rangle = \langle \cancel{O}|\hat{a}_j|\psi\rangle = \psi_j$ with $|j\rangle = \hat{a}^\dagger_j |\cancel{O}\rangle$ and $|\cancel{O}\rangle$ the vacuum state. By inserting the closure relation $\mathbb{I}=\sum_i|i\rangle \langle i|$, for $|i\rangle$ a complete basis, we find
\begin{widetext}
    \begin{equation}
    \begin{aligned}
        E\psi_{j'}=\langle j'|E | \psi \rangle = \langle j'|\hat{H} | \psi \rangle &= -t_\parallel \langle \cancel{O}| a^{}_{j'} \sum_j \left [\hat{a}^\dagger_j \hat{a}^{}_{j+1} + \hat{a}^\dagger_j \hat{a}^{}_{j-1}\right ]\sum_i|i\rangle \langle i| \psi \rangle \\
        &=-t_\parallel\sum_{j,i} \langle \cancel{O}| a^{}_{j'} \hat{a}^\dagger_j \hat{a}^{}_{j+1} |i\rangle \langle i| \psi \rangle + \langle \cancel{O}| a^{}_{j'}\hat{a}^\dagger_j \hat{a}^{}_{j-1}|i\rangle \langle i| \psi \rangle
    \end{aligned}
\end{equation}
\end{widetext}
We now observe that the orthogonality condition imposes that the only non-zero terms are for $\langle \cancel{O}| a^{}_{j'} \hat{a}^\dagger_j \hat{a}^{}_{j+1} |i\rangle=\langle \cancel{O}| a^{}_{j'} \hat{a}^\dagger_j |\cancel{O}\rangle \delta_{j+1,i} = \mathbb{I} \delta_{j',j} \delta_{j+1,i}  $ and similarly for the second term. The result reads
\begin{equation}
    E \psi_j = -t_\parallel\left ( \psi_{j+1} + \psi_{j-1}\right )
\end{equation}
By comparing with \eqref{eq:continuum_eigen}, we find that $t_\parallel = \hbar^2/2m(\Delta x)^2$. In terms of the recoil energy it now reads with $t_\parallel =E_r/\pi^2$.
Therefore, if we want to express the unitless crossover temperature from QMC in the field-theory language we have to take
\begin{equation}
    \begin{aligned}
        &\frac{T^{\text{QMC}}_{\text{cross}}}{E_r} = A \left ( \frac{t_\perp}{E_r}\right )^\nu
        \rightarrow \frac{T^{\text{QMC}}_{\text{cross}}}{t_\parallel}  = A   \pi^{2(1-\nu)} \left ( \frac{t_\perp}{t_\parallel} \right )^\nu 
    \end{aligned}
\end{equation}
where we divide and multiply by $t_\parallel$ and use that $\frac{E_r}{t_\parallel} = \pi^2$.


\begin{thebibliography}{10}
\providecommand*{\bibinfo}[2]{#2}
\providecommand*{\eprint}[1]{#1}
\providecommand*{\url}[1]{#1}
\bibitem{bloch-review-2008}
\bibinfo{author}{I.~Bloch}, \bibinfo{author}{J.~Dalibard}, and \bibinfo{author}{W.~Zwerger}, \bibinfo{title}{\emph{Many-body physics with ultracold gases}}, \bibinfo{journal}{\Jrmp} \bibinfo{volume}{\textbf{80}}, \bibinfo{pages}{885} (\bibinfo{date}{2008}).
\bibitem{giamarchi_book_1d}
\bibinfo{author}{T.~Giamarchi}, \bibinfo{title}{\emph{Quantum Physics in One Dimension}}, \bibinfo{volume}{vol. 121 of \emph{International series of monographs on physics}} (\bibinfo{publisher}{Oxford University Press}, Oxford, \bibinfo{year}{2004}).
\bibitem{hadzibabic-2Dgas-2011}
\bibinfo{author}{Z.~Hadzibabic} and \bibinfo{author}{J.~Dalibard}, \bibinfo{title}{\emph{Two-dimensional bose fluids: An atomic physics perspective}}, \bibinfo{journal}{La Rivista del Nuovo Cimento} \bibinfo{volume}{\textbf{34}}(6), \bibinfo{pages}{389} (\bibinfo{date}{2011}).
\bibitem{giamarchi_book_carr}
\bibinfo{author}{T.~Giamarchi}, in \bibinfo{editors}{L.~D. Carr}, ed., \emph{Understanding Quantum Phase Transitions} (\bibinfo{publisher}{CRC Press / Taylor \& Francis}, \bibinfo{year}{2010}), \bibinfo{pages}{p. 291}.
\bibitem{Giamarchi_2004c}
\bibinfo{author}{T.~Giamarchi}, \bibinfo{title}{\emph{Theoretical framework for quasi-one dimensional systems}}, \bibinfo{journal}{Chem. Rev.} \bibinfo{volume}{\textbf{104}}(11), \bibinfo{pages}{5037} (\bibinfo{date}{2004}).
\bibitem{bourbonnais_review_book_lebed}
\bibinfo{author}{C.~Bourbonnais} and \bibinfo{author}{D.~Jerome}, \bibinfo{title}{\emph{Interacting electrons in quasi-one-dimensional organic superconductors}} (\bibinfo{publisher}{Springer}, Heidelberg, \bibinfo{year}{2008}), \bibinfo{pages}{p. 357}, {T. Giamarchi}, ibid, p.~719.
\bibitem{klanjsek_bpcp}
\bibinfo{author}{M.~{Klanj{\v s}ek}}, \bibinfo{author}{H.~{Mayaffre}}, \bibinfo{author}{C.~{Berthier}}, \bibinfo{author}{M.~{Horvati{\'c}}}, \bibinfo{author}{B.~{Chiari}}, \bibinfo{author}{O.~{Piovesana}}, \bibinfo{author}{P.~{Bouillot}}, \bibinfo{author}{C.~{Kollath}}, \bibinfo{author}{E.~{Orignac}}, \bibinfo{author}{R.~{Citro}}, \emph{et~al.}, \bibinfo{title}{\emph{{Controlling Luttinger Liquid Physics in Spin Ladders under a Magnetic Field}}}, \bibinfo{journal}{Phys. Rev. Lett} \bibinfo{volume}{\textbf{101}}(13), \bibinfo{pages}{137207} (\bibinfo{date}{Sep. 2008}), \eprint{0804.2639}.
\bibitem{schmidiger_neutrons_bound_spinons}
\bibinfo{author}{D.~Schmidiger}, \bibinfo{author}{S.~M\"uhlbauer}, \bibinfo{author}{A.~Zheludev}, \bibinfo{author}{P.~Bouillot}, \bibinfo{author}{T.~Giamarchi}, \bibinfo{author}{C.~Kollath}, \bibinfo{author}{G.~Ehlers}, and \bibinfo{author}{A.~M. Tsvelik}, \bibinfo{title}{\emph{Symmetric and asymmetric excitations of a strong-leg quantum spin ladder}}, \bibinfo{journal}{Phys. Rev. B} \bibinfo{volume}{\textbf{88}}, \bibinfo{pages}{094411} (\bibinfo{date}{Sep 2013}).
\bibitem{schmidiger_prl2013}
\bibinfo{author}{D.~Schmidiger}, \bibinfo{author}{P.~Bouillot}, \bibinfo{author}{T.~Guidi}, \bibinfo{author}{R.~Bewley}, \bibinfo{author}{C.~Kollath}, \bibinfo{author}{T.~Giamarchi}, and \bibinfo{author}{A.~Zheludev}, \bibinfo{title}{\emph{Spectrum of a magnetized strong-leg quantum spin ladder}}, \bibinfo{journal}{Phys. Rev. Lett.} \bibinfo{volume}{\textbf{111}}, \bibinfo{pages}{107202} (\bibinfo{date}{2013}).
\bibitem{paredes_tonks_experiment}
\bibinfo{author}{B.~Paredes}, \bibinfo{author}{A.~Widera}, \bibinfo{author}{V.~Murg}, \bibinfo{author}{O.~Mandel}, \bibinfo{author}{S.~Folling}, \bibinfo{author}{I.~Cirac}, \bibinfo{author}{G.~Shlyapnikov}, \bibinfo{author}{T.~Hansch}, and \bibinfo{author}{I.~Bloch}, \bibinfo{title}{\emph{Tonks-girardeau gas of ultracold atoms in an optical lattice}}, \bibinfo{journal}{Nature} \bibinfo{volume}{\textbf{429}}, \bibinfo{pages}{277} (\bibinfo{date}{2004}).
\bibitem{Dalibard-2DBose-2005}
\bibinfo{author}{S.~Stock}, \bibinfo{author}{Z.~Hadzibabic}, \bibinfo{author}{B.~Battelier}, \bibinfo{author}{M.~Cheneau}, and \bibinfo{author}{J.~Dalibard}, \bibinfo{title}{\emph{Observation of phase defects in quasi-two-dimensional bose-einstein condensates}}, \bibinfo{journal}{\Jprl} \bibinfo{volume}{\textbf{95}}, \bibinfo{pages}{190403} (\bibinfo{date}{Nov 2005}).
\bibitem{Hofferberth2007}
\bibinfo{author}{S.~Hofferberth}, \bibinfo{author}{I.~Lesanovsky}, \bibinfo{author}{B.~Fischer}, \bibinfo{author}{T.~Schumm}, and \bibinfo{author}{J.~Schmiedmayer}, \bibinfo{title}{\emph{Non-equilibrium coherence dynamics in one-dimensional {B}ose gases}}, \bibinfo{journal}{Nature} \bibinfo{volume}{\textbf{449}}, \bibinfo{pages}{324} (\bibinfo{date}{2007}).
\bibitem{meinert-1Dexcitation-2015}
\bibinfo{author}{F.~Meinert}, \bibinfo{author}{M.~Panfil}, \bibinfo{author}{M.~J. Mark}, \bibinfo{author}{K.~Lauber}, \bibinfo{author}{J.-S. Caux}, and \bibinfo{author}{H.-C. N{\"a}gerl}, \bibinfo{title}{\emph{Probing the excitations of a {L}ieb-{L}iniger gas from weak to strong coupling}}, \bibinfo{journal}{\Jprl} \bibinfo{volume}{\textbf{115}}, \bibinfo{pages}{085301} (\bibinfo{date}{2015}).
\bibitem{kinoshita_1D_tonks_gas_observation}
\bibinfo{author}{T.~Kinoshita}, \bibinfo{author}{T.~Wenger}, and \bibinfo{author}{D.~S. Weiss}, \bibinfo{title}{\emph{Observation of a one-dimensional tonks-girardeau gas}}, \bibinfo{journal}{Science} \bibinfo{volume}{\textbf{305}}, \bibinfo{pages}{1125} (\bibinfo{date}{2004}).
\bibitem{bouchoule2011}
\bibinfo{author}{I.~Bouchoule}, \bibinfo{author}{N.~Van~Druten}, and \bibinfo{author}{C.~I. Westbrook}, \bibinfo{title}{\emph{Atom chips and one-dimensional bose gases}}, \bibinfo{journal}{Atom chips} \bibinfo{pages}{pp. 331--363} (\bibinfo{date}{2011}).
\bibitem{guo-crossoverD-2023}
\bibinfo{author}{Y.~Guo}, \bibinfo{author}{H.~Yao}, \bibinfo{author}{S.~Ramanjanappa}, \bibinfo{author}{S.~Dhar}, \bibinfo{author}{M.~Horvath}, \bibinfo{author}{L.~Pizzino}, \bibinfo{author}{T.~Giamarchi}, \bibinfo{author}{M.~Landini}, and \bibinfo{author}{H.-C. N\"agerl}, \bibinfo{title}{\emph{Observation of the 2d-1d crossover in strongly interacting ultracold bosons}}, \bibinfo{journal}{Nature Physics} \bibinfo{volume}{\textbf{20}}(6), \bibinfo{pages}{934} (\bibinfo{date}{2024}).
\bibitem{guo-cooling-2023}
\bibinfo{author}{Y.~Guo}, \bibinfo{author}{H.~Yao}, \bibinfo{author}{S.~Dhar}, \bibinfo{author}{L.~Pizzino}, \bibinfo{author}{M.~Horvath}, \bibinfo{author}{T.~Giamarchi}, \bibinfo{author}{M.~Landini}, and \bibinfo{author}{H.-C. N\"agerl}, \bibinfo{title}{\emph{Anomalous cooling of bosons by dimensional reduction}}, \bibinfo{journal}{Science Advances} \bibinfo{volume}{\textbf{10}}(7), \bibinfo{pages}{eadk6870} (\bibinfo{date}{2024}).
\bibitem{dalibard2023}
\bibinfo{author}{G.~Chauveau}, \bibinfo{author}{C.~Maury}, \bibinfo{author}{F.~Rabec}, \bibinfo{author}{C.~Heintze}, \bibinfo{author}{G.~Brochier}, \bibinfo{author}{S.~Nascimbene}, \bibinfo{author}{J.~Dalibard}, \bibinfo{author}{J.~Beugnon}, \bibinfo{author}{S.~M. Roccuzzo}, and \bibinfo{author}{S.~Stringari}, \bibinfo{title}{\emph{Superfluid fraction in an interacting spatially modulated bose-einstein condensate}}, \bibinfo{journal}{Phys. Rev. Lett.} \bibinfo{volume}{\textbf{130}}, \bibinfo{pages}{226003} (\bibinfo{date}{Jun 2023}).
\bibitem{spielman-2023}
\bibinfo{author}{J.~Tao}, \bibinfo{author}{M.~Zhao}, and \bibinfo{author}{I.~B. Spielman}, \bibinfo{title}{\emph{Observation of anisotropic superfluid density in an artificial crystal}}, \bibinfo{journal}{Phys. Rev. Lett.} \bibinfo{volume}{\textbf{131}}, \bibinfo{pages}{163401} (\bibinfo{date}{Oct 2023}).
\bibitem{pietro-2024-superfluid}
\bibinfo{author}{D.~P\'erez-Cruz}, \bibinfo{author}{G.~E. Astrakharchik}, and \bibinfo{author}{P.~Massignan}, \bibinfo{title}{\emph{Superfluid fraction of interacting bosonic gases}} (\bibinfo{date}{2024}), \eprint{arXiv:2403.08416}.
\bibitem{Jorg-hydrodynamics-crossD-2021}
\bibinfo{author}{F.~M\o{}ller}, \bibinfo{author}{C.~Li}, \bibinfo{author}{I.~Mazets}, \bibinfo{author}{H.-P. Stimming}, \bibinfo{author}{T.~Zhou}, \bibinfo{author}{Z.~Zhu}, \bibinfo{author}{X.~Chen}, and \bibinfo{author}{J.~Schmiedmayer}, \bibinfo{title}{\emph{Extension of the generalized hydrodynamics to the dimensional crossover regime}}, \bibinfo{journal}{\Jprl} \bibinfo{volume}{\textbf{126}}, \bibinfo{pages}{090602} (\bibinfo{date}{Mar 2021}).
\bibitem{biagioni-supersolid-crossD-2021}
\bibinfo{author}{G.~Biagioni}, \bibinfo{author}{N.~Antolini}, \bibinfo{author}{A.~Ala{\~n}a}, \bibinfo{author}{M.~Modugno}, \bibinfo{author}{A.~Fioretti}, \bibinfo{author}{C.~Gabbanini}, \bibinfo{author}{L.~Tanzi}, and \bibinfo{author}{G.~Modugno}, \bibinfo{title}{\emph{Dimensional crossover in the superfluid-supersolid quantum phase transition}}, \bibinfo{journal}{arXiv preprint arXiv:2111.14541}  (\bibinfo{date}{2021}).
\bibitem{cazalilla2011}
\bibinfo{author}{M.~A. Cazalilla}, \bibinfo{author}{R.~Citro}, \bibinfo{author}{T.~Giamarchi}, \bibinfo{author}{E.~Orignac}, and \bibinfo{author}{M.~Rigol}, \bibinfo{title}{\emph{One dimensional bosons: From condensed matter systems to ultracold gases}}, \bibinfo{journal}{\Jrmp} \bibinfo{volume}{\textbf{83}}, \bibinfo{pages}{1405} (\bibinfo{date}{2011}).
\bibitem{bollmark-crossoverD-2020}
\bibinfo{author}{G.~Bollmark}, \bibinfo{author}{N.~Laflorencie}, and \bibinfo{author}{A.~Kantian}, \bibinfo{title}{\emph{Dimensional crossover and phase transitions in coupled chains: Density matrix renormalization group results}}, \bibinfo{journal}{Phys. Rev. B} \bibinfo{volume}{\textbf{102}}, \bibinfo{pages}{195145} (\bibinfo{date}{Nov 2020}).
\bibitem{yao-crossD-2023}
\bibinfo{author}{H.~Yao}, \bibinfo{author}{L.~Pizzino}, and \bibinfo{author}{T.~Giamarchi}, \bibinfo{title}{\emph{{Strongly-interacting bosons at 2D-1D dimensional crossover}}}, \bibinfo{journal}{SciPost Phys.} \bibinfo{volume}{\textbf{15}}, \bibinfo{pages}{050} (\bibinfo{date}{2023}).
\bibitem{plisson-2011}
\bibinfo{author}{T.~Plisson}, \bibinfo{author}{B.~Allard}, \bibinfo{author}{M.~Holzmann}, \bibinfo{author}{G.~Salomon}, \bibinfo{author}{A.~Aspect}, \bibinfo{author}{P.~Bouyer}, and \bibinfo{author}{T.~Bourdel}, \bibinfo{title}{\emph{Coherence properties of a two-dimensional trapped bose gas around the superfluid transition}}, \bibinfo{journal}{Phys. Rev. A} \bibinfo{volume}{\textbf{84}}, \bibinfo{pages}{061606} (\bibinfo{date}{Dec 2011}).
\bibitem{cazalilla-coupled1D-2006}
\bibinfo{author}{M.~Cazalilla}, \bibinfo{author}{A.~Ho}, and \bibinfo{author}{T.~Giamarchi}, \bibinfo{title}{\emph{Interacting bose gases in quasi-one-dimensional optical lattices}}, \bibinfo{journal}{New Journal of Physics} \bibinfo{volume}{\textbf{8}}(8), \bibinfo{pages}{158} (\bibinfo{date}{2006}).
\bibitem{petrov2000a}
\bibinfo{author}{D.~S. Petrov}, \bibinfo{author}{M.~Holzmann}, and \bibinfo{author}{G.~V. Shlyapnikov}, \bibinfo{title}{\emph{{B}ose-{E}instein condensation in quasi-2d trapped gases}}, \bibinfo{journal}{\Jprl} \bibinfo{volume}{\textbf{84}}(12), \bibinfo{pages}{2551} (\bibinfo{date}{2000}).
\bibitem{petrov-2dscattering-2001}
\bibinfo{author}{D.~Petrov} and \bibinfo{author}{G.~Shlyapnikov}, \bibinfo{title}{\emph{Interatomic collisions in a tightly confined {B}ose gas}}, \bibinfo{journal}{\Jpra} \bibinfo{volume}{\textbf{64}}, \bibinfo{pages}{012706} (\bibinfo{date}{2001}).
\bibitem{haldane1981}
\bibinfo{author}{F.~D.~M. Haldane}, \bibinfo{title}{\emph{Effective harmonic-fluid approach to low-energy properties of one-dimensional quantum fluids}}, \bibinfo{journal}{\Jprl} \bibinfo{volume}{\textbf{47}}, \bibinfo{pages}{1840} (\bibinfo{date}{1981}).
\bibitem{jordan_transformation1928}
\bibinfo{author}{P.~Jordan} and \bibinfo{author}{E.~Wigner}, \bibinfo{title}{\emph{About the pauli exclusion principle}}, \bibinfo{journal}{Z. Physik} \bibinfo{volume}{\textbf{47}}, \bibinfo{pages}{631} (\bibinfo{date}{1928}).
\bibitem{cazalilla_rpm_bosons}
\bibinfo{author}{M.~A. Cazalilla}, \bibinfo{author}{R.~Citro}, \bibinfo{author}{T.~Giamarchi}, \bibinfo{author}{E.~Orignac}, and \bibinfo{author}{M.~Rigol}, \bibinfo{title}{\emph{One dimensional bosons: From condensed matter systems to ultracold gases}}, \bibinfo{journal}{Reviews of Modern Physics} \bibinfo{volume}{\textbf{83}}, \bibinfo{pages}{1405} (\bibinfo{date}{2011}).
\bibitem{Rajaraman_1982}
\bibinfo{author}{R.~Rajaraman}, \bibinfo{title}{\emph{Solitons and Instantons: An Introduction to Solitons and Instantons in Quantum Field Theory}} (\bibinfo{publisher}{North-Holland Publishing Company}, \bibinfo{year}{1982}).
\bibitem{Ho2004}
\bibinfo{author}{A.~F. Ho}, \bibinfo{author}{M.~A. Cazalilla}, and \bibinfo{author}{T.~Giamarchi}, \bibinfo{title}{\emph{Deconfinement in a 2d optical lattice of coupled 1d boson systems}}, \bibinfo{journal}{Phys. Rev. Lett.} \bibinfo{volume}{\textbf{92}}, \bibinfo{pages}{130405} (\bibinfo{date}{2004}).
\bibitem{Feynman_1972}
\bibinfo{author}{R.~Feynman}, \bibinfo{title}{\emph{Statistical Mechanics}} (\bibinfo{publisher}{Redwood City, Cal., Addison-Wesley}, \bibinfo{year}{1972}).
\bibitem{mahan2000}
\bibinfo{author}{G.~Mahan}, \bibinfo{title}{\emph{Many Particle Physics}} (\bibinfo{publisher}{Springer, New York}, \bibinfo{year}{2000}).
\bibitem{cazalilla2004_bosonizing_cold}
\bibinfo{author}{M.~A. Cazalilla}, \bibinfo{title}{\emph{Bosonizing one-dimensional cold atomic gases}}, \bibinfo{journal}{J. Phys. B} \bibinfo{volume}{\textbf{37}}, \bibinfo{pages}{S1} (\bibinfo{date}{2004}).
\bibitem{ceperley-PIMC-1995}
\bibinfo{author}{D.~M. Ceperley}, \bibinfo{title}{\emph{Path integrals in the theory of condensed helium}}, \bibinfo{journal}{\Jrmp} \bibinfo{volume}{\textbf{67}}, \bibinfo{pages}{279} (\bibinfo{date}{1995}).
\bibitem{boninsegni-worm-short-2006}
\bibinfo{author}{M.~Boninsegni}, \bibinfo{author}{N.~Prokof'ev}, and \bibinfo{author}{B.~Svistunov}, \bibinfo{title}{\emph{Worm algorithm for continuous-space path integral {M}onte {C}arlo simulations}}, \bibinfo{journal}{\Jprl} \bibinfo{volume}{\textbf{96}}, \bibinfo{pages}{070601} (\bibinfo{date}{2006}).
\bibitem{boninsegni-worm-long-2006}
\bibinfo{author}{M.~Boninsegni}, \bibinfo{author}{N.~V. Prokof'ev}, and \bibinfo{author}{B.~V. Svistunov}, \bibinfo{title}{\emph{{W}orm algorithm and diagrammatic {M}onte {C}arlo: A new approach to continuous-space path integral {M}onte {C}arlo simulations}}, \bibinfo{journal}{\Jpre} \bibinfo{volume}{\textbf{74}}, \bibinfo{pages}{036701} (\bibinfo{date}{2006}).
\bibitem{yao-boseglass-2020}
\bibinfo{author}{H.~Yao}, \bibinfo{author}{T.~Giamarchi}, and \bibinfo{author}{L.~Sanchez-Palencia}, \bibinfo{title}{\emph{{L}ieb-{L}iniger bosons in a shallow quasiperiodic potential: {B}ose glass phase and fractal {M}ott lobes}}, \bibinfo{journal}{\Jprl} \bibinfo{volume}{\textbf{125}}, \bibinfo{pages}{060401} (\bibinfo{date}{2020}).
\bibitem{gautier-2Dquasicrystal-2021}
\bibinfo{author}{R.~Gautier}, \bibinfo{author}{H.~Yao}, and \bibinfo{author}{L.~Sanchez-Palencia}, \bibinfo{title}{\emph{Strongly interacting bosons in a two-dimensional quasicrystal lattice}}, \bibinfo{journal}{\Jprl} \bibinfo{volume}{\textbf{126}}(11), \bibinfo{pages}{110401} (\bibinfo{date}{2021}).
\bibitem{cazalilla_review_bosons}
\bibinfo{author}{M.~A. Cazalilla}, \bibinfo{author}{R.~Citro}, \bibinfo{author}{T.~Giamarchi}, \bibinfo{author}{E.~Orignac}, and \bibinfo{author}{M.~Rigol}, \bibinfo{title}{\emph{One dimensional bosons: From condensed matter systems to ultracold gases}}, \bibinfo{journal}{Rev. Mod. Phys.} \bibinfo{volume}{\textbf{83}}, \bibinfo{pages}{1405} (\bibinfo{date}{Dec 2011}).
\bibitem{kantian_crossover_2023}
\bibinfo{author}{G.~Bollmark}, \bibinfo{author}{T.~K\"ohler}, \bibinfo{author}{L.~Pizzino}, \bibinfo{author}{Y.~Yang}, \bibinfo{author}{J.~S. Hofmann}, \bibinfo{author}{H.~Shi}, \bibinfo{author}{S.~Zhang}, \bibinfo{author}{T.~Giamarchi}, and \bibinfo{author}{A.~Kantian}, \bibinfo{title}{\emph{Solving 2d and 3d lattice models of correlated fermions---combining matrix product states with mean-field theory}}, \bibinfo{journal}{Phys. Rev. X} \bibinfo{volume}{\textbf{13}}, \bibinfo{pages}{011039} (\bibinfo{date}{2023}).
\bibitem{cazalilla_06}
\bibinfo{author}{M.~A. Cazalilla}, \bibinfo{title}{\emph{Effect of suddenly turning on interactions in the {Luttinger} model}}, \bibinfo{journal}{Phys. Rev. Lett.} \bibinfo{volume}{\textbf{97}}, \bibinfo{pages}{156403} (\bibinfo{date}{Oct 2006}).
\bibitem{Cheneau2012}
\bibinfo{author}{M.~Cheneau}, \bibinfo{author}{P.~Barmettler}, \bibinfo{author}{D.~Poletti}, \bibinfo{author}{M.~Endres}, \bibinfo{author}{P.~Schau\ss{}}, \bibinfo{author}{T.~Fukuhara}, \bibinfo{author}{C.~Gross}, \bibinfo{author}{I.~Bloch}, \bibinfo{author}{C.~Kollath}, and \bibinfo{author}{S.~Kuhr}, \bibinfo{title}{\emph{Light-cone-like spreading of correlations in a quantum many-body system}}, \bibinfo{journal}{\Jnature} \bibinfo{volume}{\textbf{481}}, \bibinfo{pages}{484} (\bibinfo{date}{2012}).
\bibitem{Carcy-2019-kcorrelation}
\bibinfo{author}{C.~Carcy}, \bibinfo{author}{H.~Cayla}, \bibinfo{author}{A.~Tenart}, \bibinfo{author}{A.~Aspect}, \bibinfo{author}{M.~Mancini}, and \bibinfo{author}{D.~Cl\'ement}, \bibinfo{title}{\emph{Momentum-space atom correlations in a mott insulator}}, \bibinfo{journal}{Phys. Rev. X} \bibinfo{volume}{\textbf{9}}, \bibinfo{pages}{041028} (\bibinfo{date}{Nov 2019}).
\bibitem{tenart-2021-kcorrelation}
\bibinfo{author}{A.~Tenart}, \bibinfo{author}{G.~Herc{\'e}}, \bibinfo{author}{J.-P. Bureik}, \bibinfo{author}{A.~Dareau}, and \bibinfo{author}{D.~Cl{\'e}ment}, \bibinfo{title}{\emph{Observation of pairs of atoms at opposite momenta in an equilibrium interacting bose gas}}, \bibinfo{journal}{Nature Physics} \bibinfo{volume}{\textbf{17}}(12), \bibinfo{pages}{1364} (\bibinfo{date}{2021}).
\bibitem{leggett1970}
\bibinfo{author}{A.~J. Leggett}, \bibinfo{title}{\emph{Can a solid be "superfluid"?}}, \bibinfo{journal}{Phys. Rev. Lett.} \bibinfo{volume}{\textbf{25}}, \bibinfo{pages}{1543} (\bibinfo{date}{Nov 1970}).
\bibitem{leggett1998superfluid}
\bibinfo{author}{A.~Leggett}, \bibinfo{title}{\emph{On the superfluid fraction of an arbitrary many-body system at t= 0}}, \bibinfo{journal}{Journal of statistical physics} \bibinfo{volume}{\textbf{93}}, \bibinfo{pages}{927} (\bibinfo{date}{1998}).
\bibitem{troyer1998}
\bibinfo{author}{M.~Troyer}, \bibinfo{author}{B.~Ammon}, and \bibinfo{author}{E.~Heeb}, \bibinfo{title}{\emph{Parallel object oriented {M}onte {C}arlo simulations}}, \bibinfo{journal}{Lect. Notes Comput. Sci.} \bibinfo{volume}{\textbf{1505}}, \bibinfo{pages}{191} (\bibinfo{date}{1998}).
\bibitem{ALPS2007}
\bibinfo{author}{A.~Albuquerque}, \bibinfo{author}{F.~Alet}, \bibinfo{author}{P.~Corboz}, \bibinfo{author}{P.~Dayal}, \bibinfo{author}{A.~Feiguin}, \bibinfo{author}{S.~Fuchs}, \bibinfo{author}{L.~Gamper}, \bibinfo{author}{E.~Gull}, \bibinfo{author}{S.~Guertler}, \bibinfo{author}{A.~Honecker}, \emph{et~al.}, \bibinfo{title}{\emph{The {ALPS} project release 1.3: Open-source software for strongly correlated systems}}, \bibinfo{journal}{J. Magn. Magn. Mater.} \bibinfo{volume}{\textbf{310}}, \bibinfo{pages}{1187} (\bibinfo{date}{2007}).
\bibitem{ALPS2011}
\bibinfo{author}{B.~Bauer}, \bibinfo{author}{L.~D. Carr}, \bibinfo{author}{H.~Evertz}, \bibinfo{author}{A.~Feiguin}, \bibinfo{author}{J.~Freire}, \bibinfo{author}{S.~Fuchs}, \bibinfo{author}{L.~Gamper}, \bibinfo{author}{J.~Gukelberger}, \bibinfo{author}{E.~Gull}, \bibinfo{author}{S.~Guertler}, \emph{et~al.}, \bibinfo{title}{\emph{The {ALPS} project release 2.0: Open source software for strongly correlated systems}}, \bibinfo{journal}{J. Stat. Mech.: Th. Exp.} \bibinfo{volume}{\textbf{05}}, \bibinfo{pages}{P05001} (\bibinfo{date}{2011}).

\end{thebibliography}
\end{document}